\documentclass[twocolumn,prl,showpacs,preprintnumbers,aps,altaffilletter,amsmath,superscriptaddress,amssymb,floatfix,nofootinbib,showpacs]{revtex4-1}

\usepackage[normalem]{ulem}
\usepackage{verbatim}
\usepackage{graphicx}
\usepackage{dcolumn}
\usepackage{bm}
\usepackage[usenames]{xcolor}
\usepackage{url}
\usepackage{amsmath}
\usepackage{morefloats}
\usepackage{times}
\usepackage[varg]{txfonts}
\usepackage{multirow}
\usepackage{lineno}
\usepackage{hyperref}

\usepackage{color}
\definecolor{amber}{rgb}{1.0, 0.75, 0.0}
\definecolor{orange}{rgb}{1.0, 0.5, 0.0}
\definecolor{amaranth}{rgb}{0.9, 0.17, 0.31}
\definecolor{purple}{rgb}{0.8, 0.0, 0.8}

\definecolor{amethyst}{rgb}{0.6, 0.4, 0.8}

\definecolor{burntorange}{rgb}{0.8, 0.33, 0.0}

\usepackage[normalem]{ulem}	

\usepackage{svn-multi}
\newcommand{\MONEobsCOMPACT}{\macro{\ensuremath{39_{-4}^{+6}}}} 
\newcommand{\MTWOobsCOMPACT}{\macro{\ensuremath{32_{-5}^{+4}}}} 
\newcommand{\PEMATCHSNRCOMPACT}{\macro{\ensuremath{25.3_{-0.2}^{+0.1}}}} 
\newcommand{\MFINALobsCOMPACT}{\macro{\ensuremath{68_{-4}^{+4}}}} 
\newcommand{\MFINALSCOMPACT}{\macro{\ensuremath{62_{-4}^{+4}}}} 
\newcommand{\SPINFINALCOMPACT}{\macro{\ensuremath{0.67_{-0.07}^{+0.05}}}} 

\newcommand{\macro}[1]{\textcolor{red}{#1}} 

\newcommand\OBSEVENTDATEMONTHDAYYEAR{\macro{September~14,~2015}}

\newcommand\OBSEVENTTIME{\macro{09:50:45}}

\newcommand\OBSEVENTAPPROXCOMBINEDSNR{\macro{\ensuremath{24}}}

\newcommand\CBCEVENTFAPBOUND{\macro{\ensuremath{< 2\times 10^{-7}}}}

\newcommand{\GRAVITONCOMPTONWAVELENGTH}{\macro{\ensuremath{10^{13}\,\mathrm{km}}}}

\graphicspath{{./figures/}}
\newcommand{\be}{\begin{equation}}
\newcommand{\ee}{\end{equation}}
\newcommand{\ba}{\begin{eqnarray}}
\newcommand{\ea}{\end{eqnarray}}

\def\ltsima{$\; \buildrel < \over \sim \;$}
\def\simlt{\lower.5ex\hbox{\ltsima}}
\def\gtsima{$\; \buildrel > \over \sim \;$}
\def\simgt{\lower.5ex\hbox{\gtsima}}

\def\betaprior{$[-3,3]$}
\def\alphaprior{$[-5,5]$}
\def\varphiprior{$[-20,20]$}
\def\TheEvent{GW150914}

\def\phm{\phantom{-}}
\def\dchizeromedian{$-0.1_{-0.1}^{+0.1}$}
\def\dchizerogr{$0.94$}
\def\dchionemedian{$-0.4_{-0.9}^{+0.0}$}		
\def\dchionegr{$0.94$}							
\def\dchitwomedian{$-0.35_{-0.35}^{+0.3}$}
\def\dchitwogr{$0.97$}
\def\dchithreemedian{$0.2_{-0.2}^{+0.2}$}
\def\dchithreegr{$0.04$}
\def\dchifourmedian{$-2.0_{-1.8}^{+1.6}$}
\def\dchifourgr{$0.98$}
\def\dchifivelmedian{$0.8_{-0.55}^{+0.6}$}
\def\dchifivelgr{$0.02$}
\def\dchisixmedian{$-1.5_{-1.1}^{+1.1}$}
\def\dchisixgr{$0.99$}
\def\dchisixlmedian{$8.9_{-6.8}^{+6.8}$}
\def\dchisixlgr{$0.02$}
\def\dchisevenmedian{$3.7_{-2.75}^{+2.6}$}
\def\dchisevengr{$0.02$}

\def\dbetatwomedian{$0.1_{-0.3}^{+0.4}$}
\def\dbetatwogr{$0.29$}
\def\dbetathreemedian{$0.1_{-0.3}^{+0.5}$}
\def\dbetathreegr{$0.38$}
\def\dalphatwomedian{$-0.1_{-0.4}^{+0.4}$}
\def\dalphatwogr{$0.68$}
\def\dalphathreemedian{$-0.5_{-1.5}^{+2.0}$}
\def\dalphathreegr{$0.67$}
\def\dalphafourmedian{$-0.1_{-0.6}^{+0.5}$}
\def\dalphafourgr{$0.61$}
\def\dchizeromedianall{$1.4_{-3.0}^{+3.3}$}
\def\dchizerograll{$0.21$}
\def\dchionemedianall{$-0.6_{-18.0}^{+17.7}$}	
\def\dchionegrall{$0.52$}							
\def\dchitwomedianall{$-3.2_{-15.2}^{+19.3}$}
\def\dchitwograll{$0.60$}
\def\dchithreemedianall{$2.6_{-15.7}^{+13.8}$}
\def\dchithreegrall{$0.41$}
\def\dchifourmedianall{$0.5_{-18.2}^{+17.3}$}
\def\dchifourgrall{$0.49$}
\def\dchifivelmedianall{$-1.5_{-16.3}^{+19.1}$}
\def\dchifivelgrall{$0.55$}
\def\dchisixmedianall{$-0.6_{-17.2}^{+18.2}$}
\def\dchisixgrall{$0.53$}
\def\dchisixlmedianall{$-2.4_{-15.2}^{+18.7}$}
\def\dchisixlgrall{$0.57$}
\def\dchisevenmedianall{$-3.4_{-14.8}^{+19.3}$}
\def\dchisevengrall{$0.59$}

\def\dbetatwomedianall{$0.15_{-0.5}^{+0.6}$}
\def\dbetatwograll{$0.35$}
\def\dbetathreemedianall{$-0.0_{-0.6}^{+0.8}$}
\def\dbetathreegrall{$0.56$}
\def\dalphatwomedianall{$-0.0_{-1.15}^{+1.0}$}
\def\dalphatwograll{$0.51$}
\def\dalphathreemedianall{$-0.0_{-4.4}^{+4.4}$}
\def\dalphathreegrall{$0.50$}
\def\dalphafourmedianall{$-0.0_{-1.1}^{+1.2}$}
\def\dalphafourgrall{$0.55$}

\def\dchizerologb{$\phm1.9\pm 0.1$}
\def\dchionelogb{$\phm1.3\pm 0.3$}					
\def\dchitwologb{$\phm1.2\pm 0.2$}
\def\dchithreelogb{$\phm1.2\pm 0.1$}
\def\dchifourlogb{$\phm0.3\pm 0.1$}
\def\dchifivellogb{$\phm0.7\pm 0.1$}
\def\dchisixlogb{$\phm0.4\pm 0.1$}
\def\dchisixllogb{$-0.2\pm 0.1$}
\def\dchisevenlogb{$-0.0\pm 0.2$}

\def\dbetatwologb{$\phm1.2\pm 0.1$}
\def\dbetathreelogb{$\phm0.6\pm 0.1$}
\def\dalphatwologb{$\phm1.1\pm 0.1$}
\def\dalphathreelogb{$\phm1.3\pm 0.1$}
\def\dalphafourlogb{$\phm1.2\pm 0.1$}
\def\dchislogb{$2.1\pm 0.6$}					

\def\dbetaslogb{$2.2\pm 0.1$}
\def\dalphaslogb{$2.1\pm 0.1$}

\renewcommand{\macro}[1]{{#1}}

\begin{document}

\title[\TheEvent{} and GR]{Tests of general relativity with \TheEvent{}}

\author{%
B.~P.~Abbott,$^{1}$  
R.~Abbott,$^{1}$  
T.~D.~Abbott,$^{2}$  
M.~R.~Abernathy,$^{1}$  
F.~Acernese,$^{3,4}$
K.~Ackley,$^{5}$  
C.~Adams,$^{6}$  
T.~Adams,$^{7}$
P.~Addesso,$^{3}$  
R.~X.~Adhikari,$^{1}$
% ...
\\
\medskip
(LIGO Scientific Collaboration and Virgo Collaboration) 
\\
\medskip
M.~Boyle,$^{56}$
M.~Campanelli,$^{102}$
D.~A.~Hemberger,$^{77}$
L.~E.~Kidder,$^{56}$
S.~Ossokine,$^{29}$
M.~A.~Scheel,$^{77}$
B.~Szilagyi,$^{77,130}$
S.~Teukolsky,$^{56}$\\
and Y.~Zlochower$^{102}$\\
\medskip
{{}$^{\dag}$Deceased, May 2015. {}$^{\ddag}$Deceased, March 2015. }%
}\noaffiliation
\affiliation {LIGO, California Institute of Technology, Pasadena, CA 91125, USA }
\affiliation {Louisiana State University, Baton Rouge, LA 70803, USA }
\affiliation {Universit\`a di Salerno, Fisciano, I-84084 Salerno, Italy }
\affiliation {INFN, Sezione di Napoli, Complesso Universitario di Monte S.Angelo, I-80126 Napoli, Italy }
\affiliation {University of Florida, Gainesville, FL 32611, USA }
\affiliation {LIGO Livingston Observatory, Livingston, LA 70754, USA }
\affiliation {Laboratoire d'Annecy-le-Vieux de Physique des Particules (LAPP), Universit\'e Savoie Mont Blanc, CNRS/IN2P3, F-74941 Annecy-le-Vieux, France }
\affiliation {Albert-Einstein-Institut, Max-Planck-Institut f\"ur Gravi\-ta\-tions\-physik, D-30167 Hannover, Germany }
\affiliation {Nikhef, Science Park, 1098 XG Amsterdam, Netherlands }
\affiliation {LIGO, Massachusetts Institute of Technology, Cambridge, MA 02139, USA }
\affiliation {Instituto Nacional de Pesquisas Espaciais, 12227-010 S\~{a}o Jos\'{e} dos Campos, S\~{a}o Paulo, Brazil }
\affiliation {INFN, Gran Sasso Science Institute, I-67100 L'Aquila, Italy }
\affiliation {INFN, Sezione di Roma Tor Vergata, I-00133 Roma, Italy }
\affiliation {Inter-University Centre for Astronomy and Astrophysics, Pune 411007, India }
\affiliation {International Centre for Theoretical Sciences, Tata Institute of Fundamental Research, Bangalore 560012, India }
\affiliation {University of Wisconsin-Milwaukee, Milwaukee, WI 53201, USA }
\affiliation {Leibniz Universit\"at Hannover, D-30167 Hannover, Germany }
\affiliation {Universit\`a di Pisa, I-56127 Pisa, Italy }
\affiliation {INFN, Sezione di Pisa, I-56127 Pisa, Italy }
\affiliation {Australian National University, Canberra, Australian Capital Territory 0200, Australia }
\affiliation {The University of Mississippi, University, MS 38677, USA }
\affiliation {California State University Fullerton, Fullerton, CA 92831, USA }
\affiliation {LAL, Universit\'e Paris-Sud, CNRS/IN2P3, Universit\'e Paris-Saclay, 91400 Orsay, France }
\affiliation {Chennai Mathematical Institute, Chennai 603103, India }
\affiliation {Universit\`a di Roma Tor Vergata, I-00133 Roma, Italy }
\affiliation {University of Southampton, Southampton SO17 1BJ, United Kingdom }
\affiliation {Universit\"at Hamburg, D-22761 Hamburg, Germany }
\affiliation {INFN, Sezione di Roma, I-00185 Roma, Italy }
\affiliation {Albert-Einstein-Institut, Max-Planck-Institut f\"ur Gravitations\-physik, D-14476 Potsdam-Golm, Germany }
\affiliation {APC, AstroParticule et Cosmologie, Universit\'e Paris Diderot, CNRS/IN2P3, CEA/Irfu, Observatoire de Paris, Sorbonne Paris Cit\'e, F-75205 Paris Cedex 13, France }
\affiliation {Montana State University, Bozeman, MT 59717, USA }
\affiliation {Universit\`a di Perugia, I-06123 Perugia, Italy }
\affiliation {INFN, Sezione di Perugia, I-06123 Perugia, Italy }
\affiliation {European Gravitational Observatory (EGO), I-56021 Cascina, Pisa, Italy }
\affiliation {Syracuse University, Syracuse, NY 13244, USA }
\affiliation {SUPA, University of Glasgow, Glasgow G12 8QQ, United Kingdom }
\affiliation {LIGO Hanford Observatory, Richland, WA 99352, USA }
\affiliation {Wigner RCP, RMKI, H-1121 Budapest, Konkoly Thege Mikl\'os \'ut 29-33, Hungary }
\affiliation {Columbia University, New York, NY 10027, USA }
\affiliation {Stanford University, Stanford, CA 94305, USA }
\affiliation {Universit\`a di Padova, Dipartimento di Fisica e Astronomia, I-35131 Padova, Italy }
\affiliation {INFN, Sezione di Padova, I-35131 Padova, Italy }
\affiliation {CAMK-PAN, 00-716 Warsaw, Poland }
\affiliation {University of Birmingham, Birmingham B15 2TT, United Kingdom }
\affiliation {Universit\`a degli Studi di Genova, I-16146 Genova, Italy }
\affiliation {INFN, Sezione di Genova, I-16146 Genova, Italy }
\affiliation {RRCAT, Indore MP 452013, India }
\affiliation {Faculty of Physics, Lomonosov Moscow State University, Moscow 119991, Russia }
\affiliation {SUPA, University of the West of Scotland, Paisley PA1 2BE, United Kingdom }
\affiliation {University of Western Australia, Crawley, Western Australia 6009, Australia }
\affiliation {Department of Astrophysics/IMAPP, Radboud University Nijmegen, P.O. Box 9010, 6500 GL Nijmegen, Netherlands }
\affiliation {Artemis, Universit\'e C\^ote d'Azur, CNRS, Observatoire C\^ote d'Azur, CS 34229, Nice cedex 4, France }
\affiliation {MTA E\"otv\"os University, ``Lendulet'' Astrophysics Research Group, Budapest 1117, Hungary }
\affiliation {Institut de Physique de Rennes, CNRS, Universit\'e de Rennes 1, F-35042 Rennes, France }
\affiliation {Washington State University, Pullman, WA 99164, USA }
\affiliation {Cornell University, Ithaca, NY 14853, USA }
\affiliation {Universit\`a degli Studi di Urbino ``Carlo Bo,'' I-61029 Urbino, Italy }
\affiliation {INFN, Sezione di Firenze, I-50019 Sesto Fiorentino, Firenze, Italy }
\affiliation {University of Oregon, Eugene, OR 97403, USA }
\affiliation {Laboratoire Kastler Brossel, UPMC-Sorbonne Universit\'es, CNRS, ENS-PSL Research University, Coll\`ege de France, F-75005 Paris, France }
\affiliation {Astronomical Observatory Warsaw University, 00-478 Warsaw, Poland }
\affiliation {VU University Amsterdam, 1081 HV Amsterdam, Netherlands }
\affiliation {University of Maryland, College Park, MD 20742, USA }
\affiliation {Center for Relativistic Astrophysics and School of Physics, Georgia Institute of Technology, Atlanta, GA 30332, USA }
\affiliation {Institut Lumi\`{e}re Mati\`{e}re, Universit\'{e} de Lyon, Universit\'{e} Claude Bernard Lyon 1, UMR CNRS 5306, 69622 Villeurbanne, France }
\affiliation {Laboratoire des Mat\'eriaux Avanc\'es (LMA), IN2P3/CNRS, Universit\'e de Lyon, F-69622 Villeurbanne, Lyon, France }
\affiliation {Universitat de les Illes Balears, IAC3---IEEC, E-07122 Palma de Mallorca, Spain }
\affiliation {Universit\`a di Napoli ``Federico II,'' Complesso Universitario di Monte S.Angelo, I-80126 Napoli, Italy }
\affiliation {NASA/Goddard Space Flight Center, Greenbelt, MD 20771, USA }
\affiliation {Canadian Institute for Theoretical Astrophysics, University of Toronto, Toronto, Ontario M5S 3H8, Canada }
\affiliation {Tsinghua University, Beijing 100084, China }
\affiliation {Texas Tech University, Lubbock, TX 79409, USA }
\affiliation {The Pennsylvania State University, University Park, PA 16802, USA }
\affiliation {National Tsing Hua University, Hsinchu City, 30013 Taiwan, Republic of China }
\affiliation {Charles Sturt University, Wagga Wagga, New South Wales 2678, Australia }
\affiliation {University of Chicago, Chicago, IL 60637, USA }
\affiliation {Caltech CaRT, Pasadena, CA 91125, USA }
\affiliation {Korea Institute of Science and Technology Information, Daejeon 305-806, Korea }
\affiliation {Carleton College, Northfield, MN 55057, USA }
\affiliation {Universit\`a di Roma ``La Sapienza,'' I-00185 Roma, Italy }
\affiliation {University of Brussels, Brussels 1050, Belgium }
\affiliation {Sonoma State University, Rohnert Park, CA 94928, USA }
\affiliation {Northwestern University, Evanston, IL 60208, USA }
\affiliation {University of Minnesota, Minneapolis, MN 55455, USA }
\affiliation {The University of Melbourne, Parkville, Victoria 3010, Australia }
\affiliation {The University of Texas Rio Grande Valley, Brownsville, TX 78520, USA }
\affiliation {The University of Sheffield, Sheffield S10 2TN, United Kingdom }
\affiliation {University of Sannio at Benevento, I-82100 Benevento, Italy and INFN, Sezione di Napoli, I-80100 Napoli, Italy }
\affiliation {Montclair State University, Montclair, NJ 07043, USA }
\affiliation {Universit\`a di Trento, Dipartimento di Fisica, I-38123 Povo, Trento, Italy }
\affiliation {INFN, Trento Institute for Fundamental Physics and Applications, I-38123 Povo, Trento, Italy }
\affiliation {Cardiff University, Cardiff CF24 3AA, United Kingdom }
\affiliation {National Astronomical Observatory of Japan, 2-21-1 Osawa, Mitaka, Tokyo 181-8588, Japan }
\affiliation {School of Mathematics, University of Edinburgh, Edinburgh EH9 3FD, United Kingdom }
\affiliation {Indian Institute of Technology, Gandhinagar Ahmedabad Gujarat 382424, India }
\affiliation {Institute for Plasma Research, Bhat, Gandhinagar 382428, India }
\affiliation {University of Szeged, D\'om t\'er 9, Szeged 6720, Hungary }
\affiliation {Embry-Riddle Aeronautical University, Prescott, AZ 86301, USA }
\affiliation {University of Michigan, Ann Arbor, MI 48109, USA }
\affiliation {Tata Institute of Fundamental Research, Mumbai 400005, India }
\affiliation {American University, Washington, D.C. 20016, USA }
\affiliation {Rochester Institute of Technology, Rochester, NY 14623, USA }
\affiliation {University of Massachusetts-Amherst, Amherst, MA 01003, USA }
\affiliation {University of Adelaide, Adelaide, South Australia 5005, Australia }
\affiliation {West Virginia University, Morgantown, WV 26506, USA }
\affiliation {University of Bia{\l }ystok, 15-424 Bia{\l }ystok, Poland }
\affiliation {SUPA, University of Strathclyde, Glasgow G1 1XQ, United Kingdom }
\affiliation {IISER-TVM, CET Campus, Trivandrum Kerala 695016, India }
\affiliation {Institute of Applied Physics, Nizhny Novgorod, 603950, Russia }
\affiliation {Pusan National University, Busan 609-735, Korea }
\affiliation {Hanyang University, Seoul 133-791, Korea }
\affiliation {NCBJ, 05-400 \'Swierk-Otwock, Poland }
\affiliation {IM-PAN, 00-956 Warsaw, Poland }
\affiliation {Monash University, Victoria 3800, Australia }
\affiliation {Seoul National University, Seoul 151-742, Korea }
\affiliation {University of Alabama in Huntsville, Huntsville, AL 35899, USA }
\affiliation {ESPCI, CNRS, F-75005 Paris, France }
\affiliation {Universit\`a di Camerino, Dipartimento di Fisica, I-62032 Camerino, Italy }
\affiliation {Southern University and A\&M College, Baton Rouge, LA 70813, USA }
\affiliation {College of William and Mary, Williamsburg, VA 23187, USA }
\affiliation {Instituto de F\'\i sica Te\'orica, University Estadual Paulista/ICTP South American Institute for Fundamental Research, S\~ao Paulo SP 01140-070, Brazil }
\affiliation {University of Cambridge, Cambridge CB2 1TN, United Kingdom }
\affiliation {IISER-Kolkata, Mohanpur, West Bengal 741252, India }
\affiliation {Rutherford Appleton Laboratory, HSIC, Chilton, Didcot, Oxon OX11 0QX, United Kingdom }
\affiliation {Whitman College, 345 Boyer Avenue, Walla Walla, WA 99362 USA }
\affiliation {National Institute for Mathematical Sciences, Daejeon 305-390, Korea }
\affiliation {Hobart and William Smith Colleges, Geneva, NY 14456, USA }
\affiliation {Janusz Gil Institute of Astronomy, University of Zielona G\'ora, 65-265 Zielona G\'ora, Poland }
\affiliation {Andrews University, Berrien Springs, MI 49104, USA }
\affiliation {Caltech JPL, Pasadena, CA 91109, USA }
\affiliation {Universit\`a di Siena, I-53100 Siena, Italy }
\affiliation {Trinity University, San Antonio, TX 78212, USA }
\affiliation {University of Washington, Seattle, WA 98195, USA }
\affiliation {Kenyon College, Gambier, OH 43022, USA }
\affiliation {Abilene Christian University, Abilene, TX 79699, USA }

\begin{abstract}
The LIGO detection of \TheEvent{} provides an unprecedented opportunity to study the two-body motion 
of a compact-object binary in the large velocity, highly nonlinear regime, and to witness the final merger 
of the binary and the excitation of uniquely relativistic modes of the gravitational field.
We carry out several investigations to determine whether 
\TheEvent{} is consistent with a binary black-hole merger in general relativity.
We find that the final remnant's mass and spin, as determined from the low-frequency (inspiral)
and high-frequency (post-inspiral) phases of the signal, are mutually consistent with
the binary black-hole solution in general relativity. Furthermore, the data following the peak   
of \TheEvent{} are consistent with the least-damped quasi-normal mode inferred 
from the mass and spin of the remnant black hole. By using waveform models
that allow for parameterized general-relativity violations during the inspiral and
merger phases, we perform quantitative tests on the gravitational-wave phase in the 
dynamical regime and we determine the first empirical bounds on several high-order post-Newtonian 
coefficients. We constrain the graviton Compton wavelength, assuming that gravitons
are dispersed in vacuum in the same way as particles with mass, obtaining   
a $90\%$-confidence lower bound of \GRAVITONCOMPTONWAVELENGTH{}. In conclusion, within our
statistical uncertainties, we find no
evidence for violations of general relativity in the genuinely strong-field regime of gravity.
\end{abstract}

\date{\today}

\maketitle

\paragraph{Introduction.}
On \OBSEVENTDATEMONTHDAYYEAR{}, at \OBSEVENTTIME{} Universal Time,
the LIGO detectors at Hanford, Washington and Livingston,
Louisiana, detected a gravitational-wave (GW)
signal, henceforth \TheEvent{}, with an observed signal-to-noise
  ratio (SNR) $\sim$ \OBSEVENTAPPROXCOMBINEDSNR. The probability that \TheEvent{} was
  due to a random noise fluctuation was later established to be \CBCEVENTFAPBOUND{}~\cite{GW150914-DETECTION,GW150914-CBC}.
\TheEvent{} exhibited the expected signature of an inspiral, merger,  
and ringdown signal from a coalescing binary system \cite{GW150914-DETECTION}. 
Assuming that general relativity (GR) is the correct 
description for \TheEvent{}, detailed follow-up analyses determined
the (detector-frame) component 
masses of the binary system to be \MONEobsCOMPACT{} $M_\odot$ and \MTWOobsCOMPACT{} $M_\odot$ 
at 90\% credible intervals \cite{GW150914-PARAMESTIM}, corroborating the hypothesis that 
\TheEvent{} was emitted by a binary black hole.

In Newtonian gravity, binary systems move along circular or elliptical orbits with 
constant orbital period~\cite{Kepler:1609,Newton:1687}.
In GR, binary systems emit GWs~\cite{Einstein:1916,Einstein:1918}; as a consequence, the binary's orbital period decreases over time 
as energy and angular momentum are radiated away. Electromagnetic observations of binary
pulsars over the four decades since their discovery
\cite{Hulse1975Discovery,Taylor:1982} have made it possible to measure
GW-induced orbital-period variations $\dot{P}_{\rm orb} \sim -
10^{-14} \mbox{--} 10^{-12}$, confirming the GW luminosity
predicted at leading order in post-Newtonian (PN) theory~\cite{Blanchet:2014} 
(i.e., Einstein's quadrupole formula) with exquisite precision~\cite{Burgay:2003jj,Wex:2014nva}.
Nevertheless, even in the most relativistic binary
pulsar known today, J0737-3039~\cite{Burgay:2003jj}, the orbital period changes at an effectively constant 
rate. The orbital velocity $v$ relative to the speed of light $c$ is $v/c \sim 2 \times 10^{-3}$, 
and the two neutron stars in the system will coalesce in $\sim 85\,{\rm Myr}$.

By contrast, GW150914 was emitted by a rapidly evolving, dynamical
binary that swept through the detectors' bandwidth and merged in a fraction of a second, with $\dot{P}_{\rm orb}$ ranging from $\sim -0.1$ at $f_{\rm GW} \sim 30\ {\rm Hz}$ to $\sim -1$ at $f_{\rm GW} \sim 132\ {\rm Hz}$ (just before merger, where $v/c$ reached $\sim 0.5$).
Thus, through GW150914 
we observe the two-body motion in the large-velocity, highly
dynamical, strong-field regime of gravity, leading to the formation of
a new merged object, and generating GWs. While Solar-System
experiments, binary-pulsar observations, and cosmological measurements
are all in excellent agreement with GR (see
Refs.~\cite{Will:2014kxa,Wex:2014nva,Berti:2015itd} and references
therein),  they test it in low-velocity, quasi-static, weak-field, or
linear regimes.\footnote{While the orbits of binary pulsars are weakly relativistic, pulsars themselves are strongly self-gravitating bodies, so they do offer opportunities to test strong-field gravity~\cite{Damour:1992we,Freire:2012mg}.}
Thus, GW150914 opens up the distinct opportunity of probing unexplored sectors of GR.

Here we perform several studies of \TheEvent{},
aimed at detecting deviations from the predictions of GR.
Within the limits set by LIGO's sensitivity and by the nature of \TheEvent{},
we find no statistically significant evidence against the
hypothesis that \TheEvent{} was emitted by two black holes spiraling
towards each other and merging to form a single, rotating black hole 
\cite{Schwarzschild:1916uq,Kerr:1963ud}, and that 
the dynamics of the process as a whole was in accordance with the vacuum Einstein field 
equations.


We begin by constraining the level of coherent (i.e., GW-like) residual strain left 
after removing the most-probable GR waveform from the \TheEvent{} data, and
use this estimated level to bound GR violations which are not degenerate with
changes in the parameters of the binary. We then verify that the mass and spin 
parameters of the final black hole, as predicted 
from the binary's inspiral signal, are consistent with the final parameters
inferred from the post-inspiral (merger and ringdown) signal.
We find that the data following the peak of \TheEvent{} are consistent with the 
least-damped quasi-normal mode (QNM) inferred from the final black-hole's characteristics. 
Next, we perform targeted measurements of the PN and phenomenological
coefficients that parameterize theoretical waveform 
models, and find no tension with the values predicted in
GR and numerical-relativity (NR) simulations. Furthermore, 
we search for evidence of dispersion in the propagation of \TheEvent{} 
toward the Earth, as it would appear in a theory in which the graviton 
is assigned a finite Compton wavelength (i.e., a nonzero mass). Finally, we show that, due to the LIGO network configuration, we cannot exclude the presence of non-GR polarization states in \TheEvent{}.
\begin{figure}
\begin{center}
\includegraphics[width=\columnwidth]{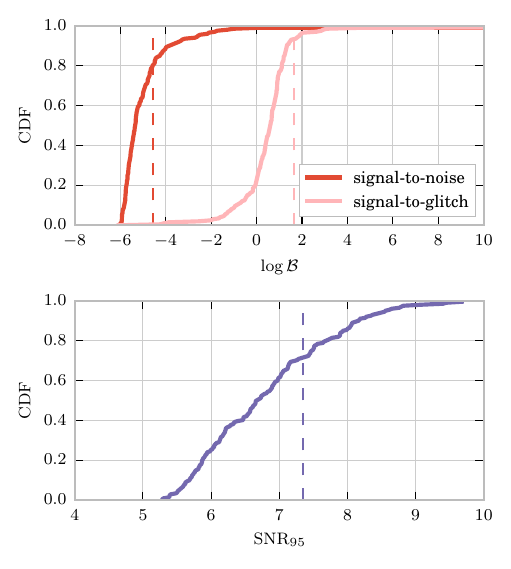}
\end{center}
\vspace*{-0.2in}
\caption{Upper panel: cumulative distribution function (CDF) of $\log$ Bayes factor -- 
the logarithm of the ratio of Bayesian evidences between two competing models --
for the signal-versus-noise 
and signal-versus-glitch 
\textsc{BayesWave} models, computed for 100 4-s stretches of data around \TheEvent{}. 
Lower Panel: cumulative distribution function (CDF) of the 95\%
credible upper bound on network coherent-burst SNR, denoted SNR$_{95}$, again computed for
100 instrument-noise segments. In both panels, we indicate with dashed
lines the $\log$ Bayes factors and upper bound on  coherent-burst
  SNR corresponding to the residuals obtained after subtracting the most probable waveform from \TheEvent{}.}
\label{fig:residual}
\end{figure}

As we shall see, the constraints on the strong-field dynamics of gravity obtained from
\TheEvent{} are not yet very tight; for instance, some of the bounds on relative deviations 
in PN parameters are $\mathcal{O}(1)$. On the other hand, it is to be noted that the LIGO detectors are
still a factor of a few away from their final design sensitivities \cite{Aasi:2013wya}, and
even louder sources than \TheEvent{} may be seen in the near future; 
moreover, as more detections are made, we will be able to combine information from all 
observed sources to obtain progressively sharper bounds on PN and other coefficients.

In the rest of this paper, when reporting physical quantities that are redshifted in 
the transformation between the source and detector frames, we refer to the detector 
frame unless we specify otherwise.

\paragraph{Waveform models, systematics, and statistical effects.} Tests of GR 
from GW observations build on the knowledge of the gravitational waveform in GR, and on 
the statistical properties of  instrumental noise. Any uncontrolled
systematic effect from waveform modeling and/or the 
detectors could in principle affect the outcome of our tests. Thus, we 
begin by checking that these uncertainties are either below our measurement 
precision or accounted for.

The analytical inspiral-merger-ringdown (IMR) waveform models used in this paper 
were developed within two frameworks: i) the effective-one-body (EOB) formalism
\cite{BuonannoDamour:1999,BuonannoDamour:2000,Damour:2008qf,Damour:2009kr,Barausse:2009xi},
which combines PN results~\cite{Blanchet:2014} with
NR~\cite{Pretorius:2005gq,Campanelli:2005dd,Baker:2005vv} and
perturbation theory~\cite{Vishveshwara:1970zz,Press:1971wr,
  Chandrasekhar:1975zza}, and ii) a phenomenological
approach~\cite{Pan:2007nw,Ajith:2007kx,Ajith:2009bn,Santamaria:2010yb}
based on extending frequency-domain PN expressions and hybridizing
PN/EOB with NR waveforms. In particular, here we adopt the double-spin,
nonprecessing waveform model developed in
Ref.~\cite{Taracchini:2013rva} using NR waveforms from
Ref.~\cite{Mroue:2013xna}, enhanced with reduced-order modeling~\cite{field2014fast} to
speed up waveform generation \cite{Purrer:2014fza,Purrer:2015tud}
(henceforth, \textsc{EOBNR}), and the single-effective--spin,
precessing waveform model of
Refs.~\cite{Husa:2015iqa,Khan:2015jqa,Hannam:2013oca} (henceforth,
\textsc{IMRPhenom}).\footnote{The specific names of the two waveform
  models that we use in the \textsc{LIGO Algorithm Library} are
  \textsc{SEOBNRv2\_ROM\_DoubleSpin} and \textsc{IMRPhenomPv2}.} Both
models are calibrated against waveforms from direct numerical
integration of the Einstein equations.  

As shown in Refs.~\cite{Taracchini:2013rva,Khan:2015jqa,Kumar:2016dhh,GW150914-PARAMESTIM,GW150914-ACCURACY},
in the region of parameter space relevant for \TheEvent{}, the error 
due to differences between the two analytical waveform models (and between the 
analytical and numerical-relativity waveforms) is smaller than the typical statistical uncertainty due to the 
finite SNR of \TheEvent{}. To assess potential modeling systematics, we collected existing NR waveforms and
generated new, targeted simulations. The simulations were generated with multiple independent 
codes \cite{Bruegmann:2006at,O'Shaughnessy:2012ay,Scheel:2014ina,
Chu:2015kft,Lousto:2015uwa,Szilagyi:2015rwa}, and sample the posterior region 
for the masses and spins inferred for GW150914 
\cite{GW150914-PARAMESTIM}. 
Since the posteriors for the magnitudes and orientations of the component spins 
are not very constraining, the choices for these parameters covered wide ranges. 
To validate the studies below, we added the publicly available and
new NR waveforms as mock signals to the data in the neighbourhood of \TheEvent{}~\cite{NRinj,Mroue:2013xna,Szilagyi:2015rwa}.
A further possible cause for systematics are uncertainties in the
calibration of the gravitational-strain observable in the LIGO detectors. These uncertainties are
modeled and included in the results presented here 
according to the treatment detailed in Ref.~\cite{GW150914-PARAMESTIM}.

\paragraph{Residuals after subtracting the most-probable waveform model.} 
The burst analysis~\cite{GW150914-BURST}, which looks for unmodeled transients and hence 
does not rely on theoretical signal templates, 
can be used to test the consistency of \TheEvent{} with
waveform models derived from GR. Using the \textsc{LALInference}~\cite{Veitch:2014wba} Bayesian-inference software library, we 
identify the most probable (i.e., \textit{maximum a posteriori}, henceforth MAP) binary black-hole waveform~\cite{GW150914-PARAMESTIM}, compute its effect in the Livingston and Hanford
detectors, and then subtract it from the data. If the data are consistent with the theoretical 
signal, no detectable power should remain after subtraction other than what is consistent with instrumental noise.
We analyze the residual with the \textsc{BayesWave}~\cite{Cornish:2014kda} algorithm developed to characterize generic GW transients.
\textsc{BayesWave} uses the evidence ratio (Bayes factor) to
rank competing hypotheses given the observed data. We compare predictions
from models in which:  (i) the data contain only Gaussian noise; (ii)
the data contain Gaussian noise and uncorrelated noise transients, or
glitches, and (iii) the data contain Gaussian noise and an
elliptically polarized GW signal. We compute the signal-to-noise Bayes factor, which is a measure of significance for the excess power in the data, and the signal-to-glitch Bayes factor, which measures the coherence of the excess power between the two detectors. 

Our analysis reveals that the \TheEvent{} residual favors the instrumental noise hypothesis over the 
presence of a coherent signal as well as the presence of glitches in either detectors; see the dashed 
lines in the top panel of Fig.~1. The positive Bayes factor 
for the signal-to-glitch hypotheses indicates that the data prefer the presence of a coherent signal
over glitches; nevertheless, the signal remains below common significance thresholds, as indicated by the limit on the residual 
$\mathrm{SNR}_{\mathrm{res}}$ given in the lower panel of Fig.~\ref{fig:residual} and further 
explained below. This is an indication of the stability of the 
LIGO detectors at the time of \TheEvent{}.
We also apply the same analysis to 100 4-second long segments of data drawn within a few minutes of 
\TheEvent{}, and
produce the cumulative distribution functions of Bayes factors shown in the upper panel of Fig.~\ref{fig:residual}. 
We find that, according to the burst analysis, the GW150914 residual is not statistically distinguishable from the instrumental noise recorded in the vicinity of the
detection, suggesting that all of the measured power is well represented by the GR prediction for the signal from a binary 
black-hole merger. The results of this analysis are very similar regardless of the MAP waveform used (i.e., \textsc{EOBNR} or \textsc{IMRPhenom}).

\begin{figure}
\includegraphics[width=0.95\columnwidth]{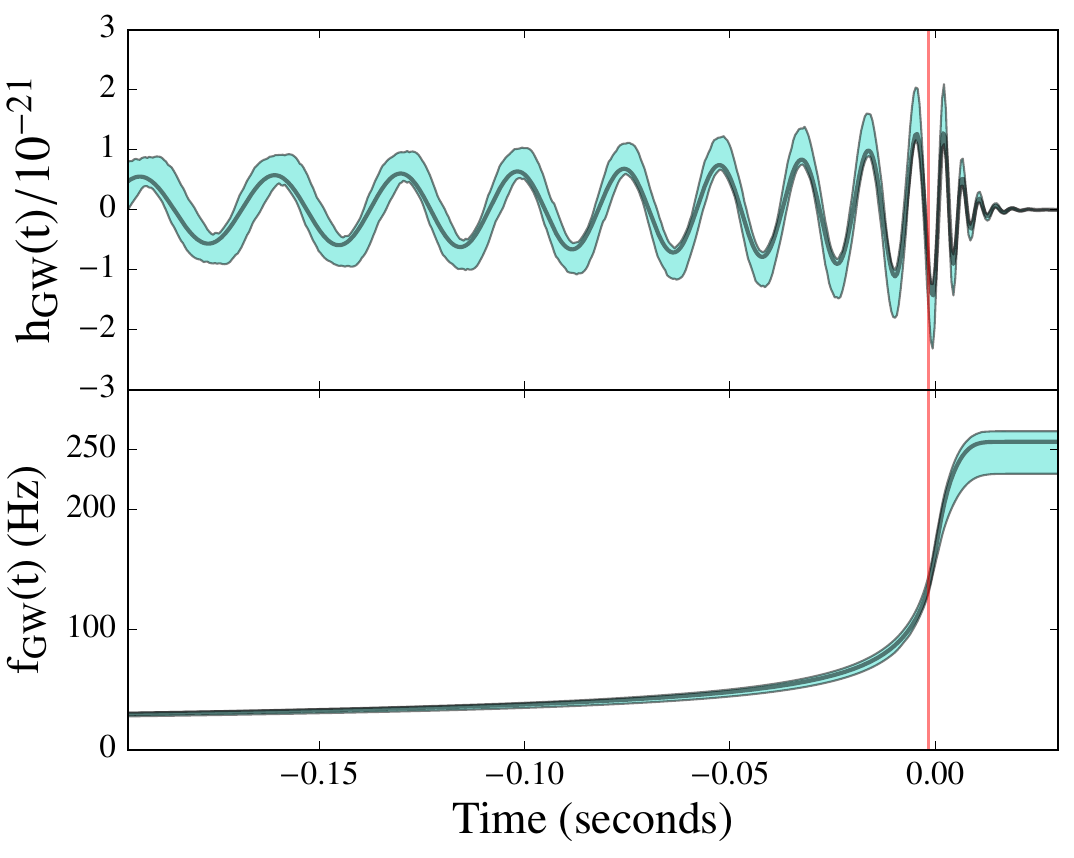}
\caption{MAP estimate and 90\% credible regions for the waveform (upper panel) and GW frequency (lower panel) of 
\TheEvent{} as estimated by the \textsc{LALInference}  analysis~\cite{GW150914-PARAMESTIM}. The solid 
lines in each panel indicate the most probable waveform from \TheEvent{}~\cite{GW150914-PARAMESTIM} and its GW frequency. 
We mark with a vertical line the instantaneous frequency $f_{\rm GW}^{\rm end \,insp} = 132$ Hz, which is used
in the IMR consistency test to delineate the boundary between the frequency-domain inspiral and post-inspiral parts 
(see Fig.~\ref{fig:regions} below for a representation of the most probable waveform's amplitude in frequency domain).}
\label{fig:imr_waveform}
\end{figure}

\begin{figure}[t]
\includegraphics[width=\columnwidth]{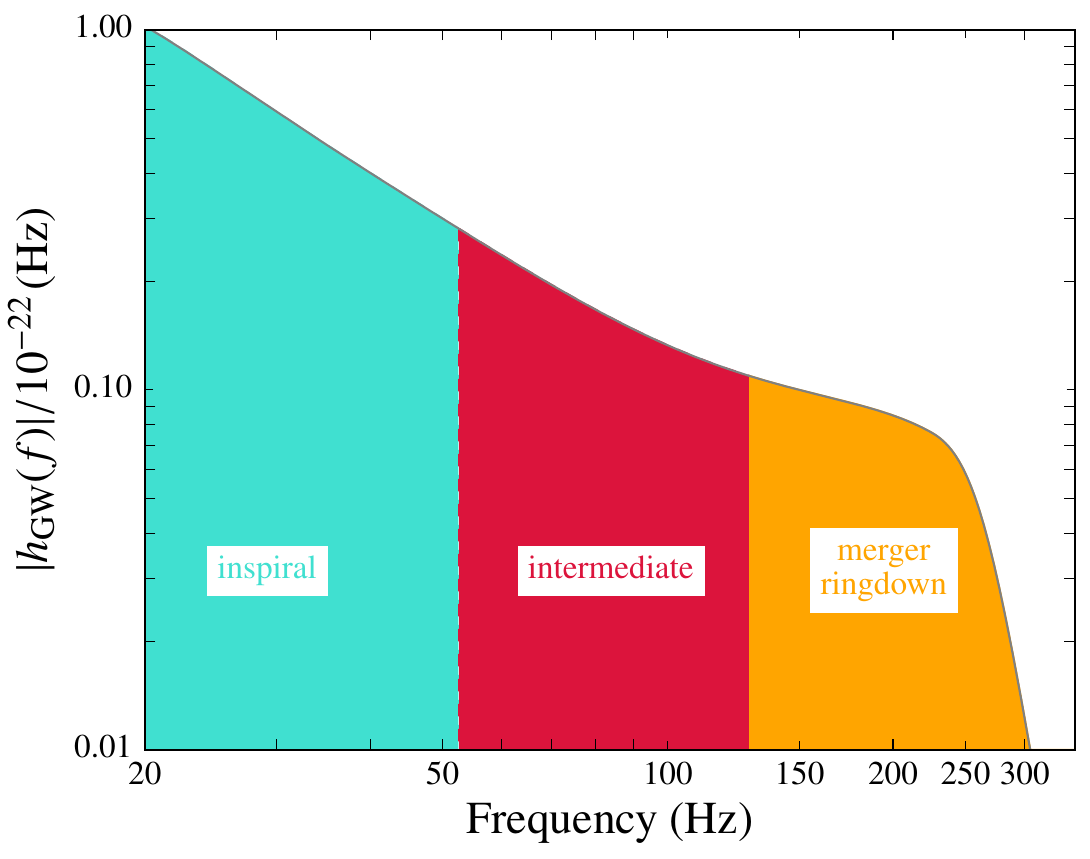}
\caption{Frequency regions of the parameterized waveform model as defined in the text and in
  Ref.~\cite{Khan:2015jqa}. The plot shows the absolute value of the frequency-domain amplitude of the 
most-probable waveform from \TheEvent{}~\cite{GW150914-PARAMESTIM}. The inspiral region (cyan) from 20\,Hz 
to $\sim$55\,Hz corresponds to the early and late inspiral regimes. The intermediate region (red) 
goes from $\sim$ 55\,Hz to $\sim$ 130\,Hz. Finally, the merger--ringdown region (orange) goes 
from $\sim$ 130\,Hz to the end of the waveform.} 
\label{fig:regions}
\end{figure}

We compute the 95\% upper bound on the coherent network ${\rm SNR}_\mathrm{res}$. This upper bound is $\mathrm{SNR}_{\mathrm{res}} \leq 7.3$ at 95\% confidence, independently of the MAP waveform used (i.e.,
\textsc{EOBNR} or \textsc{IMRPhenom}). We note that this coherent-burst SNR has a different meaning compared to the (modeled) matched-filtering binary-coalescence SNR of 24 cited for \TheEvent{}. Indeed, the upper-limit $\mathrm{SNR}_{\mathrm{res}}$ inferred for \TheEvent{} lies in the typical range for the data segments around
  \TheEvent{} (see the bottom panel of Fig.~\ref{fig:residual}), so it can be attributed to instrument noise alone.

If we assume that ${\rm SNR}_{\mathrm{res}}$ is entirely due to the mismatch between the MAP waveform 
and the underlying true signal, and 
that the putative violation of GR cannot be reabsorbed in the waveform model 
by biasing the estimates of the physical parameters~\cite{VallisneriYunes:2013,VitaleDelPozzo:2014}, we can constrain
the minimum \textit{fitting factor} ($\mathrm{FF}$)
\cite{Apostolatos:1995} between the MAP model and \TheEvent{}. 
An imperfect fit to the data leaves $\mathrm{SNR}_{\mathrm{res}}^2 = (1-\mathrm{FF}^2)\,\mathrm{FF}^{-2}\,\mathrm{SNR}_{\rm det}^2$ \cite{CornishEtAl:2011,Vallisneri:2012} where $\mathrm{SNR_{\rm det}} =$\PEMATCHSNRCOMPACT{} is the network SNR inferred by \textsc{LALInference}~\cite{GW150914-PARAMESTIM}. 
$\mathrm{SNR}_{\mathrm{res}} \leq 7.3$ then implies $\mathrm{FF} \geq 0.96$. Considering that, for parameters similar to those inferred for \TheEvent, our waveform models have much higher FFs against numerical GR waveforms, we conclude that the noise-weighted
correlation between the observed strain signal and the true GR waveform is $\geq 96\%$. 
This statement can be read as implying that the GR prediction for \TheEvent{} is verified to 
better than 4\%, in a precise sense related to noise-weighted signal correlation; and conversely, that effects due to GR-violations
in \TheEvent{} are limited to less than 4\% (for effects that cannot be reabsorbed in a redefinition of physical parameters).

\begin{figure}
\includegraphics[width=0.95\columnwidth]{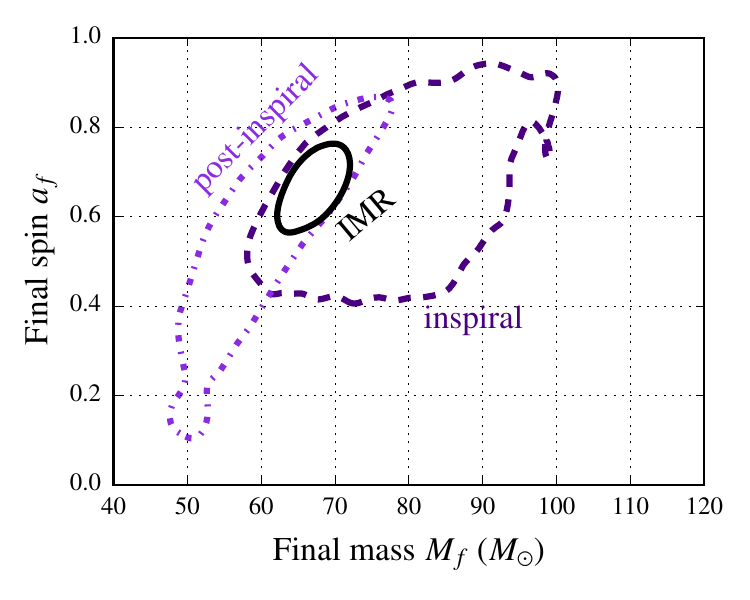}\\
\vspace{0.1in}
\includegraphics[width=0.95\columnwidth]{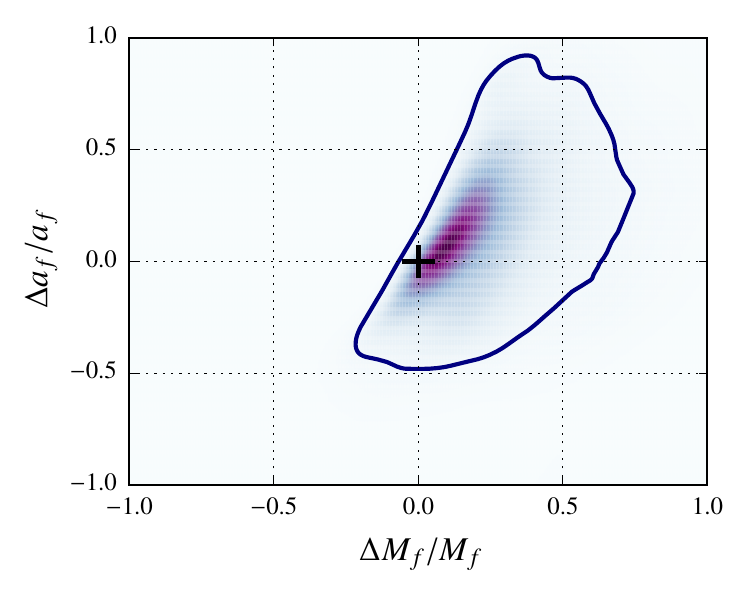}
\caption{\textit{Top panel}: 90\% credible
    regions in the joint posterior distributions for the mass $M_f$
    and dimensionless spin $a_f$ of the final compact object as determined from the
    inspiral (dark violet, dashed) and post-inspiral (violet, 
dot-dashed) signals, and from a full
    inspiral--merger--ringdown analysis (black). \textit{Bottom panel}: Posterior 
    distributions for the parameters
    $\Delta M_f/M_f$ and $\Delta a_f/a_f$ that describe the
    fractional difference in the estimates of the final mass and spin from inspiral and
    post-inspiral signals. The contour shows the $90\%$
    confidence region. The plus symbol indicates the
    expected GR value $(0,0)$.}
\label{fig:imr_consistency_test}
\end{figure}

\paragraph{Inspiral--merger--ringdown consistency test.} We now perform a test to show that the entire \TheEvent{} waveform does not deviate from the predictions of a
binary black-hole coalescence in GR. One way to do that is to compare the estimates of the mass and
spin of the remnant obtained from the low-frequency and high-frequency parts of the waveform, 
using the relations between the binary's components and final masses and spins provided by 
NR~\cite{Healy:2014yta}.

For the purpose of this test, we choose $f^{\rm end \,insp}_{\rm GW}=132$ Hz as the frequency at which the late inspiral
phase ends. In Fig.~\ref{fig:imr_waveform} we plot the EOBNR MAP waveform~\cite{GW150914-PARAMESTIM} and its 90\% credible intervals, as well as the corresponding instantaneous frequency; the vertical line marks $f^{\rm end \,insp}$. Fig.~\ref{fig:regions}
shows the frequency-domain MAP waveform amplitude; note that 132 Hz lies just before what is generally denoted as the merger--ringdown phase in the frequency domain.

To perform the test, we first truncate the frequency-domain representation of the waveforms to lie between $20$ Hz to $f^{\rm end \,insp}_{\rm GW}$, and we estimate the posterior distributions of the binary's component masses and spins using this ``inspiral'' (low-frequency) part of the observed signal,
using the nested-sampling algorithm in the \textsc{LALInference} software library~\cite{Veitch:2014wba}. We then use formulae obtained from NR simulations to compute posterior distributions of the remnant's mass and spin.
Next, we obtain the complementary ``post-inspiral'' (high-frequency) signal, which is dominated by the contribution from the merger and ringdown stages, by restricting the frequency-domain representation of the waveforms to extend between 
$f^{\rm end\,insp}_{\rm GW}$ and $1024$ Hz. Again, we derive the posterior distributions of the component masses and spins, and (by way of NR-derived formulae) of the mass and spin of the final compact object. We note that the
  MAP waveform has an expected ${\rm SNR}_{\rm det} \sim 19.5$ if we truncate
  its frequency-domain representation to have support between 20 and 132 Hz,
  and $\sim 16$ if we truncate it to have support between 132 and 1,024 Hz.
Finally, we
compare these
two estimates of the final $M_f$ and dimensionless spin $a_f$, and compare them also against the estimate performed using full inspiral--merger--ringdown waveforms.
In all cases, we average the posteriors obtained with the \textsc{EOBNR} and \textsc{IMRPhenom} waveform models, following the procedure outlined in Ref.~\cite{GW150914-PARAMESTIM}.
Technical details about the implementation of this test can be found in Ref.~\cite{Ghosh:2015xx}. 

This test is similar in spirit to the $\chi^2$ GW search statistic~\cite{Allen:2004gu,GW150914-CBC},
which divides the model waveform into frequency bands and checks that
the SNR accumulates as expected across those bands.  Large matched-filter SNR values which are accompanied by large $\chi^2$ statistic
are very likely due either to noise glitches, or to a mismatch between
the signal and the model matched-filter waveform.  Conversely, reduced-$\chi^2$ values near unity indicate that the data are consistent
with waveform plus the expected detector noise.  Thus, large
$\chi^2$ values are a warning that some parts of
the waveform are fit much worse than others, and thus the candidates 
may be due to instrument glitches that are very loud, but do not resemble
binary-inspiral signals. However, $\chi^2$ tests are performed 
by comparing the data with a single theoretical waveform, while in
this case we allow the inspiral and post-inspiral 
partial waveforms to select different physical parameters. Thus, this
test should be sensitive to subtler deviations from the predictions of GR.

In Fig.~\ref{fig:imr_consistency_test} 
we summarize our findings. The top panel shows 
the posterior distributions of $M_f$ and $a_f$ estimated from the inspiral and post-inspiral signals, 
and from the entire inspiral--merger--ringdown waveform. The plot confirms the expected behavior: 
the inspiral and post-inspiral $90\%$ confidence regions (defined by the isoprobability contours 
that enclose $90\%$ of the posterior) have a significant region of overlap. As a sanity check
(which strictly speaking is not part of the test of GR that is being performed) we also produced the 90\% 
confidence region computed with the full inspiral-merger-ringdown waveform; it lies comfortably
within this overlap.
We have verified that these conclusions are not affected by the specific formula~\cite{Healy:2014yta,Pan:2011gk,
Husa:2015iqa} used to predict $M_f$ and $a_f$, nor by the choice of $f_{\rm GW}^{\rm end\,insp}$ within $\pm 50$ Hz.

\begin{figure}[t]
\includegraphics[width=0.9\columnwidth]{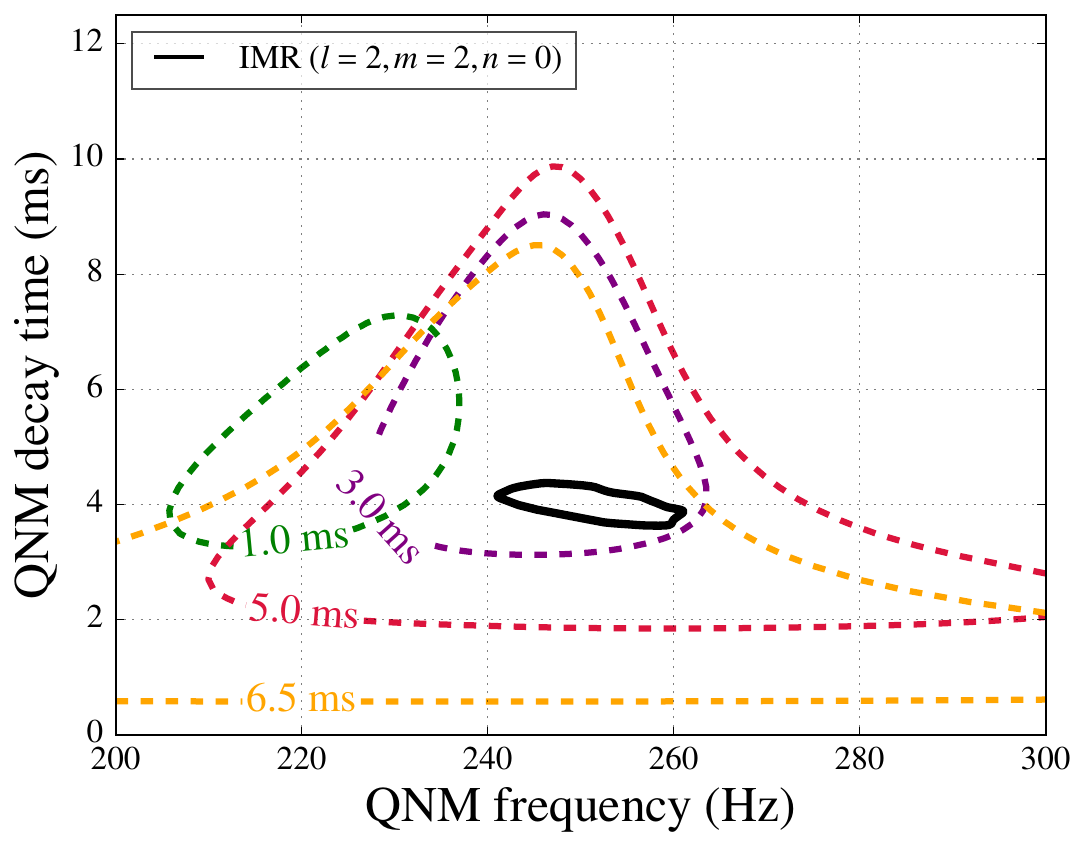}
\caption{$90\%$ credible regions in the joint posterior distributions for the damped-sinusoid parameters $f_0$ and $\tau$ (see main text), assuming start times $t_0 = t_M + 1,3,5,6.5$ ms, where $t_M$ is the merger time of the MAP waveform for \TheEvent{}. The black solid line shows the $90\%$ credible region for the frequency and decay time of
 the $\ell =2$, $m=2$, $n=0$ (i.e., the least damped) QNM, as derived from the posterior distributions of the remnant mass and spin parameters.}
\label{fig:QNM}
\end{figure}

To assess the significance of our findings more quantitatively, we define parameters $\Delta
M_f/M_f$ and $\Delta a_f/a_f$ that describe the fractional
difference between the two estimates of the final mass and spin, and calculate
their joint posterior distribution, using for $(M_f, a_f)$ the posterior distribution
obtained from the full IMR waveform; see \cite{Ghosh:2015xx} for explicit expressions. 
The result is shown in the bottom panel of
Fig.~\ref{fig:imr_consistency_test}; 
the solid line marks the isoprobability contour that contains $90\%$ of the posterior.
The plus symbol indicates the null $(0,0)$ result expected in GR, which lies on the 
isoprobability contour that encloses $28\%$ of the posterior.

We have checked that if we perform this analysis on NR signals added to LIGO instrumental noise, the null 
$(0,0)$ result expected in GR lies within the iso-probability
contour that encloses $68\%$ of the posterior roughly 68\% of
the time, as expected from random noise fluctuations.
By contrast, our test can rule out the null hypothesis (with high statistical 
significance) when analyzing a simulated signal that reflects a significant GR violation in
the frequency dependence of the energy and angular momentum
loss~\cite{Ghosh:2015xx}, even when we choose violations which would be too small to 
be noticeable in double-pulsar observations~\cite{Wex:2014nva}; for an explicit 
example we refer to Fig.~1 of Ref.~\cite{Ghosh:2015xx}. This includes signals 
with $\chi^2$ value close to unity, so that they would not have been missed by
the modeled-signal searches. 
Thus, our inspiral--merger--ringdown test shows no evidence of discrepancies with the 
predictions of GR. 

The component masses and spins estimated in Ref.~\cite{GW150914-PARAMESTIM}, together with NR-derived relations, imply $M_f =
\MFINALobsCOMPACT{} M_\odot$ (\MFINALSCOMPACT{} $M_\odot$ in the source
frame) and $a_f = \SPINFINALCOMPACT{}$ at $90\%$ confidence. 
From the posterior distributions of the mass and spin of the final black hole, we can predict the 
frequency and decay time of the least-damped QNM (i.e., the $\ell=2,m=2, n=0$ overtone)
~\cite{Berti:2005ys}. We find $f^{\rm QNM}_{220}=251_{-8}^{+8}$ Hz 
and $\tau^{\rm QNM}_{220}=4.0_{-0.3}^{+0.3}$ ms at $90\%$ confidence. 

\paragraph{Testing for the least-damped QNM in the data.}
We perform a test to check the consistency of the data with the predicted least-damped QNM of the remnant black hole.
For this purpose we compute the Bayes factor between a damped-sinusoid waveform model and
Gaussian noise, and estimate the corresponding parameter posteriors.
The signal model used is $h(t\ge t_0) = A\,e^{-(t-t_0)/\tau}\,\cos\left[2\pi\,f_0\,(t-t_0) + \phi_0\right]$,
$h(t<t_0)=0$, with fixed starting time $t_0$, and uniform
priors over the unknown frequency $f_0\in[200,300]\,$Hz and damping time 
$\tau\in[0.5, 20]\,$ms. The prior on amplitude $A$ and phase $\phi_0$ is chosen as a
two-dimensional Gaussian isotropic prior in
$\{A_s\equiv -A\sin\phi_0,\,A_c\equiv A\cos\phi_0\}$ with a
characteristic scale $H$, which is in turn marginalized over the range
$H\in[2,10]\times10^{-22}$ with a prior $\propto 1/H$.
This is a practical choice that encodes relative ignorance 
about the detectable damped-sinusoid amplitude in this range.
We use $8\,$s of data (centered on \TheEvent) from both
detectors, band-passed to ${[20,1900]}\,$Hz.
The data are analyzed coherently, assuming the signal arrived $7\,$ms
earlier at Livingston compared to Hanford, and the amplitude received in 
the two detectors has approximately equal magnitude and opposite sign (as seen in
e.g.~Fig.~1 of \cite{GW150914-DETECTION}).

We compute the Bayes factor and posterior estimates of $\{f_0,\tau\}$ as a
function of the unknown QNM start-time $t_0$, which we parameterize as an offset 
from a fiducial GPS merger time\footnote{The merger time is obtained by taking the \textsc{EOBNR} MAP waveform and
lining this waveform up with the data such that the largest SNR is obtained.
The merger time is then defined as the point at which the quadrature sum of
the $h_+$ and $h_{\times}$ polarizations is maximum.} 
$t_{\mathrm{M}} =$ 1,126,259,462.423 s (at the LIGO Hanford site).
Figure~\ref{fig:QNM} shows the $90\%$ credible contours in
the $\{f_0,\tau\}$ plane as a function of the merger-to-start time offset $t_0-t_{\mathrm{M}}$, as well as the corresponding contour for the least-damped QNM as predicted in GR for the remnant mass and spin parameters estimated for \TheEvent{}.

The $90\%$ posterior contour starts to overlap with GR prediction from
the IMR waveform for $t_0 = t_{\mathrm{M}} + 3\,$ms, or $\sim 10\,M$ after merger.
The corresponding log Bayes factor at this point is $\log_{10}B \sim 14$ and the MAP waveform SNR is $\sim 8.5$.
For $t_0 = t_{\mathrm M} + 5\,$ms the MAP parameters fall within the contour predicted
in GR for the least-damped QNM, with
$\log_{10}B\sim 6.5$ and ${\rm SNR} \sim 6.3$.
At $t_0 =t_{\mathrm M} + 6.5\,$ms, or about $20\,M$ after
merger, the Bayes factor is $\log_{10}B\sim 3.5$ with ${\rm SNR} \sim 4.8$.
The signal becomes undetectable shortly thereafter, for
$t_0 \gtrsim t_{\mathrm M} + 9\,$ms, where $B\lesssim1$.

Measuring the frequency and decay time of \textit{one} damped
sinusoid in the data does not by itself allow us to conclude that we have
observed the least-damped QNM of the final black hole, since 
the measured quality factor could be biased by the presence of the
other QNMs in the ringdown signal (see, e.g., Ref.~\cite{Dreyer:2003bv,Berti:2005ys} 
and references therein).
However, based on the numerical simulations discussed in 
Refs.~\cite{Buonanno:2006ui,Berti:2007fi,Kamaretsos:2011um}, 
one should expect the GW frequency to level off at 
$10-20\, M$ after the merger, which is where the description of ringdown in 
terms of QNMs becomes valid. For a mass $M \sim 68\,M_\odot$, the 
corresponding range is $\sim 3-7$ ms after merger. Since this is where we observe 
the 90\% posterior contours of the damped-sinusoid waveform model  
and the 90\% confidence region estimated from the IMR waveform to be 
consistent with each other, we may conclude that the data are 
compatible with the presence of the least-damped QNM as predicted by GR.  

In the future, we will extend the analysis to two damped sinusoids, and explore 
the possibility of independently extracting the final black hole's
mass and spin. A test of the general relativistic no-hair theorem~\cite{Israel:1967wq,Carter:1971zc} requires
the identification of at least two QNM frequencies in the ringdown
waveform~\cite{Dreyer:2003bv,Gossan:2011ha,Meidam:2014jpa}. 
Such a test would benefit from the observation of a system with 
a total mass similar to the one of \TheEvent{}, but with a larger asymmetry between 
component masses, which would increase the amplitudes of the sub-dominant 
modes; a stronger misalignment of the orbital angular momentum
with the line of sight would further improve their visibility 
\cite{Gossan:2011ha}. 
Finally, the determination of the remnant mass and spin independently of binary
 component parameters will allow us to test the second law 
of black-hole dynamics~\cite{Hawking:1971tu,Bardeen:1973gs}.

\paragraph{Constraining parameterized deviations from general-relativistic inspiral--merger--ringdown waveforms.} 
Because \TheEvent{} was emitted by a binary black hole in its final phase of rapid orbital evolution, 
its gravitational phasing (or phase evolution) encodes nonlinear conservative and dissipative effects that are not observable in binary pulsars, 
whose orbital period changes at an approximately constant 
rate.\footnote{Current binary-pulsar observations do constrain conservative dynamics at 1PN order
and they partially constrain spin--orbit effects at 1.5PN order through geodetic spin precession~\cite{Wex:2014nva}.}
Those effects include tails of radiation due to backscattering of GWs by the curved background around the 
coalescing black holes \cite{Blanchet:1993ec}, nonlinear tails (i.e., tails of tails) \cite{Blanchet:1997jj}, 
couplings between black-hole spins and the binary's orbital angular momentum, interactions between 
the spins of the two bodies~\cite{LenseThirring:1918,BarkerOConnell:1975,Kidder:1995zr}, and excitations of 
QNMs~\cite{Vishveshwara:1970zz,Press:1971wr,Chandrasekhar:1975zza} as the remnant black hole settles in the stationary configuration.

Whether all these subtle effects can actually be identified in \TheEvent{} and tested against GR predictions 
depends of course on their strength with respect to instrument noise and on whether the available waveform models are 
parameterized in terms of those physical effects. \TheEvent{} is moderately loud, 
with ${\rm SNR} \sim \OBSEVENTAPPROXCOMBINEDSNR{}$, certainly much smaller than what can be achieved in binary-pulsar
observations. Our ability to analyze the fine structure of the 
\TheEvent{} waveform is correspondingly limited. Our approach is to adopt a parameterized analytical family 
of inspiral--merger--ringdown waveforms, then treat the waveform coefficients as free variables that can be estimated 
(either individually or in groups) from the \TheEvent{} data~\cite{Blanchet:1994,Blanchet:1995,Blanchet:1995aa,Arun:2006hn,Mishra:2010tp,YP09,
Li:2011cg}. We can then verify that the posterior probability distributions for the coefficients include their GR values. 

The simplest and fastest parameterized waveform model that is currently available~\cite{Khan:2015jqa} can be used to bound physical effects only for the coefficients that 
enter the early inspiral phase, because for the late inspiral, merger, and ringdown phases it uses phenomenological 
coefficients fitted to NR waveforms. Louder GW events, to be collected as detector sensitivity improves, and more sophisticated 
parameterized waveform models, will allow us to do much more stringent and physical tests targeted at specific relativistic effects. We work within a subset of the TIGER framework~\cite{Li:2011cg,AgathosEtAl:2014} and 
perform a null-hypothesis test by comparing \TheEvent{} with a \textit{generalized}, analytical inspiral--merger--ringdown waveform model 
(henceforth, \textsc{gIMR}) that includes parameterized deformations with 
respect to GR. In this framework, deviations from GR are modeled as
{\it fractional changes} $\{\delta \hat{p}_i\}$ in any of the
parameters $\{p_i\}$ that parameterize the GW phase expression 
in the baseline waveform model. Similarly to Refs.~\cite{Li:2011cg,AgathosEtAl:2014}, we only 
consider deviations from GR in the GW phase, while we leave the GW amplitude unperturbed. Indeed, at the 
SNR of \TheEvent{} (i.e., ${\rm SNR} \sim \OBSEVENTAPPROXCOMBINEDSNR{}$), we expect to have much higher sensitivity to the GW 
phase rather than to its amplitude. Also, amplitude deviations
  could be reabsorbed in the calibration error model used to analyze \TheEvent{}~\cite{GW150914-PARAMESTIM}.

\begin{figure}[t]
\includegraphics[width=\columnwidth]{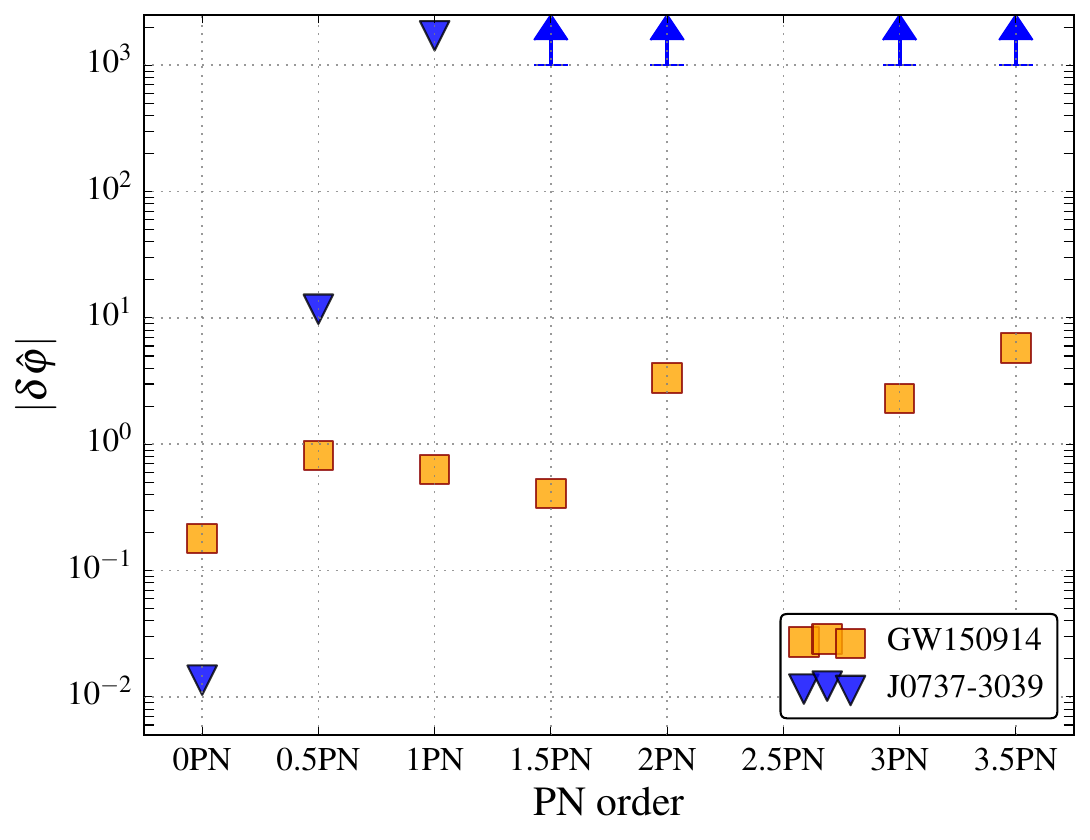}
\caption{$90\%$ upper bounds on the fractional variations of the known
  PN coefficients with respect to their GR values. The
  orange squares are the 90\% upper bounds obtained from the
  single-parameter analysis of \TheEvent{}. As a comparison, the blue triangles
show the $90\%$ upper bounds extrapolated exclusively from the measured orbital-period 
derivative $\dot{P}_{\rm orb}$ of the double pulsar J0737-3039~\cite{YunesHughes2010,Wex:2014nva},
here too allowing for possible GR violations at different powers of frequency, 
one at a time. 
The GW phase deduced from an almost constant $\dot{P}_{\rm orb}$ cannot provide 
significant information as the PN order is increased, so we show the 
bounds for the latter only up to 1PN order. We do not 
report on the deviation of the 2.5PN coefficient, which is unmeasurable because it is degenerate with the reference 
phase. We also do not report on the deviations of the logarithmic terms in the PN series at 2.5PN and 3PN order, which can
be found in Table~\ref{tab:tiger-parameters} and in Fig.~~\ref{fig:summary}.} 
\label{fig:PNbounds}
\end{figure}

\begin{figure*}[t]
\includegraphics[width=\textwidth]{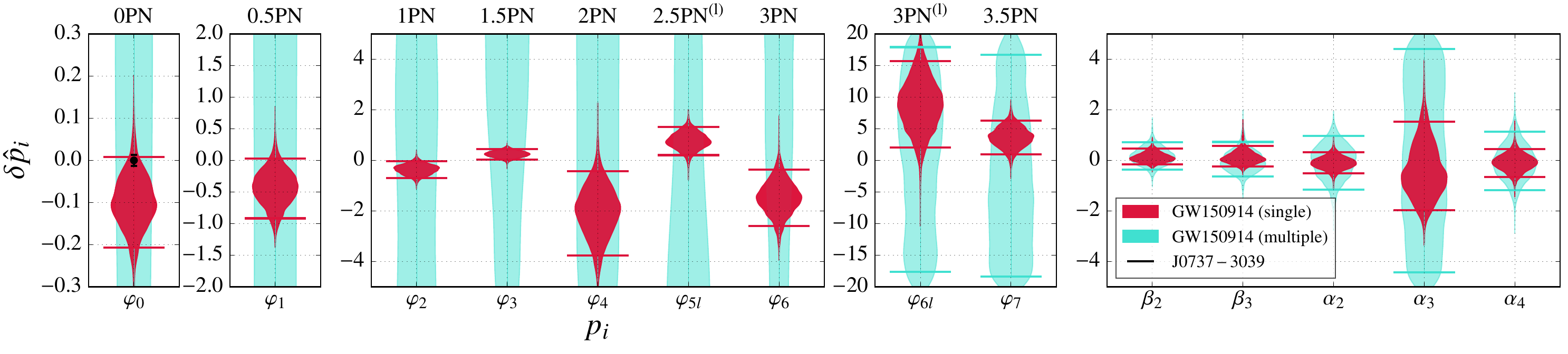}
\caption{{\it Violin} plot summarizing the posterior probability
  density distributions for all the
parameters in the \textsc{gIMR} model. (Summary statistics are reported in
Table~\ref{tab:tiger-parameters}.) From left to right: the
  plot shows increasingly high-frequency regimes as outlined in the
  text and Fig.~\ref{fig:regions}; the leftmost posteriors, labeled from 0PN to 3.5PN, are for the early-inspiral
  PN regime; the $\beta_i$ and $\alpha_i$ parameters 
  correspond to the intermediate and merger--ringdown regimes. Note that the constraints get tighter in the merger and 
ringdown regimes.
In red, we show posterior
probability distributions for the single-parameter analysis, while in cyan
we show the posterior distribution for the multiple-parameter analysis. The black
error bar at 0PN shows the bound inferred from the double pulsar; higher PN orders 
are not shown as their constraints are far weaker than \TheEvent{}'s measurement and they would appear
in the plot as vertical black lines covering the entire $y$-axis. The 2.5PN term reported in the figure refers to the logarithmic term
$\delta\hat{\varphi}_{5l}$. Because of their very different scale
compared to the rest of the parameters, the 0PN and 0.5PN posterior
distributions from \TheEvent{} and the double-pulsar limits at 0PN order 
are shown on separate panels. The error bars indicate the
90\% credible regions reported in Table~\ref{tab:tiger-parameters}
(their placement is corrected in this version). 
Due to correlations among parameters, the
  posterior distribution obtained from the multiple-parameter 
  analyses in the early-inspiral regimes are un-informative. }
\label{fig:summary}
\end{figure*}

We construct \textsc{gIMR} starting from the frequency-domain \textsc{IMRPhenom} waveform model. 
The dynamical stages that characterize the coalescence process can be represented in the 
frequency-domain by plotting the absolute value of the waveform's amplitude. We review those stages 
in Fig.~\ref{fig:regions} to guide the reader towards the interpretation of the
  results that are summarized in Table~\ref{tab:tiger-parameters} and
  Figs.~\ref{fig:PNbounds} and \ref{fig:summary}. We refer to the
  {\it early-inspiral stage} as the PN part of the
  GW phase. This stage of the phase evolution is known analytically up to
  $(v/c)^7$ and it is parameterized in terms of the PN coefficients
  $\varphi_j$, $j=0,\ldots,7$ and the \textit{logarithmic} terms 
  $\varphi_{jl}$,
  $j=5,6$. The {\it late-inspiral stage}, parameterized in terms of
  $\sigma_j$, $j=1,\ldots,4$, is defined as the phenomenological
  extension of the PN series to $(v/c)^{11}$. The \textit{early} and \textit{late}
    \textit{inspiral} stages are denoted simply as \textit{inspiral} both in
    Ref.~\cite{Khan:2015jqa} and in Fig.~\ref{fig:regions}. The {\it
    intermediate stage} that models the transition between the
  inspiral and the merger--ringdown phase is parameterized in terms of the phenomenological coefficients 
$\beta_j$,  $j=1,2,3$. Finally, the {\it merger--ringdown} phase is
  parameterized in terms of the phenomenological coefficients $\alpha_j$, $j=1,2,3$. 
The $\beta_j$ and $\alpha_j$ aim to capture the frequency dependences of the phase of the 
  corresponding regimes; see the column ``$f$--dependence" in Table \ref{tab:tiger-parameters}.  
    Due to the procedure through which the model is
  constructed, which involves fitting a waveform phasing ansatz to a calibration set of EOB waveforms 
  joined to NR 
waveforms~\cite{Khan:2015jqa}, there is an intrinsic uncertainty in the values of the 
  phenomenological parameters of the \textsc{IMRPhenom}
  model. For the intermediate and merger--ringdown regime, we verified that these intrinsic 
  uncertainties are much smaller than
  the corresponding statistical uncertainties for \TheEvent{}, and thus
do not affect our conclusions. 
In the late-inspiral case, the uncertainties associated with the calibration
of the $\sigma_j$ parameters are large, and almost comparable with
the statistical measurement uncertainties. For this reason, we do not report results for the 
$\sigma_j$ parameters.

As said, we construct the \textsc{gIMR} model by introducing (fractional) deformations $\delta\hat{p}_i$ 
for each of the \textsc{IMRPhenom}
phase parameters $p_i$, which dominate the evolution of the phase at the different stages in the
coalescence explained above. At each point in parameter space, the coefficients $p_i$ are evaluated 
for the local physical parameters (masses, spins) and multiplied by factors $(1 + \delta \hat{p}_i)$.
When using such waveforms as templates, the parameters that are allowed to vary freely are 
then the ones that are also present in the GR waveforms (masses, spins, sky position, orientation, distance,
and a reference time and phase), together with one or more of the $\delta\hat{p}_i$; the 
$p_i$ themselves are calculated using their GR expressions in terms of masses and spins.  
In this parameterization, GR is uniquely defined as the locus in the parameter space where all of
the testing parameters $\delta \hat{p}_i$ are zero. In summary, our battery of testing parameters consists of:
(i) early-inspiral stage: $\{\delta\hat{\varphi}_0,\delta\hat{\varphi}_1, \delta\hat{\varphi}_2,\delta\hat{\varphi}_3, 
§\delta\hat{\varphi}_4, \delta\hat{\varphi}_{5l}, \delta\hat{\varphi}_6,$ 
$\delta\hat{\varphi}_{6l}, \delta\hat{\varphi}_7\}$\footnote{Unlike Ref.~\cite{Khan:2015jqa}, we explicitly include the
logarithmic terms $\delta\hat{\varphi}_{5l}$ and $\delta\hat{\varphi}_{6l}$. We also 
include the 0.5PN parameter $\delta\hat{\varphi}_1$; since $\varphi_1$ is zero in GR, we define
$\delta\hat{\varphi}_1$ to be an absolute shift rather than a fractional deformation.}, 
(ii) intermediate regime: $\{\delta\hat{\beta}_2, \delta\hat{\beta}_3\}$, 
and (iii) merger--ringdown regime: $\{\delta\hat{\alpha}_2, \delta\hat{\alpha}_3, \delta\hat{\alpha}_4\}$.
We do not consider parameters that are degenerate with either the
reference time or the reference phase. For our analysis, we explore
two scenarios: {\it single-parameter} analysis, in which only one of
the testing parameters is allowed to vary freely (in addition to masses, spins, ...) while 
the remaining ones are fixed to their GR value, that is zero, 
and {\it multiple-parameter} analysis in which all the parameters in one of the     
three sets enumerated above are allowed to vary simultaneously.

The rationale behind our choices of single- and multiple-parameter analyses comes from the following considerations. 
In most known alternative theories of gravity~\cite{Will:2014kxa,Berti:2015itd,Yunes:2013dva}, the corrections to GR
extend to all PN orders even if in most cases they have been computed only at leading PN order.
Considering that GW150914 is an inspiral--merger--ringdown signal sweeping through the detector 
between 20 Hz and 300 Hz, we expect to see signal deviations from GR at
all PN orders. The single-parameter analysis corresponds to minimally extended models that can capture
  deviations from GR that occur predominantly, but not only, at a
  specific PN order. Nevertheless, should a deviation be measurably present at multiple PN 
  orders, we expect the single-parameter analyses to also capture these. In the multiple-parameter
  analysis, the correlations among the parameters are very
  significant. In other words, a shift in one of the
    testing parameters can always be compensated by a change of the opposite sign in another parameter, and still return the same overall GW phase. Thus, it
    is not surprising that the multiple-parameter case
    provides a much more conservative statement on the agreement between
\TheEvent{} and GR. We defer to future studies the identification of optimally determined directions in the $\delta\hat{p}_i$ space by performing a singular value decomposition along the lines suggested in Ref.~\cite{Pai:2012mv}.

\begin{table*}[t]
\centering
\caption{Summary of results for the \textsc{gIMR} parameterized-deviation analysis of \TheEvent{}. For each parameter
  in the \textsc{gIMR} model, we report its frequency dependence, its median and 90\% credible intervals, the quantile of the GR value of 0 in the 1D posterior probability density function. Finally,
the last two columns show $\log_{10}$ Bayes
factors between GR and the \textsc{gIMR} model. The
uncertainties on the log Bayes factors are $2\sigma$. The $a$ and $b$ coefficients shown for $\delta\hat{\alpha}_4$ are functions of the component masses and
spins (see Ref.~\cite{Khan:2015jqa}). For each field, we report the corresponding
quantities for both the single-parameter and multiple-parameter analyses.}
\label{tab:tiger-parameters}
\vspace{6pt}
\begin{tabular}{l|ccccccccc}
\hline\hline
\multicolumn{1}{c|}{waveform regime} & & &
\multicolumn{2}{c}{median}& \multicolumn{2}{c}{GR quantile} & \multicolumn{2}{c}{$\log_{10} B^{\rm GR}_{\rm model}$}\\
&parameter&$f-$dependence&single&multiple&single&multiple&single&multiple\\
\hline\hline
\multirow{9}{*}{early-inspiral regime}& $\delta\hat{\varphi}_0$  &
$f^{-5/3}$ &  \dchizeromedian  & \dchizeromedianall      &\dchizerogr
& \dchizerograll   & \dchizerologb  &\multirow{9}{*}{\dchislogb }\\   
&$\delta\hat{\varphi}_1$  & $f^{-4/3}$ &  \dchionemedian &\dchionemedianall     &\dchionegr & \dchionegrall   &\dchionelogb & \\  
&$\delta\hat{\varphi}_2$  & $f^{-1}$ &  \dchitwomedian &\dchitwomedianall     &\dchitwogr & \dchitwograll   & \dchitwologb &  \\  
&$\delta\hat{\varphi}_3$  & $f^{-2/3}$ &  \dchithreemedian
&\dchithreemedianall      & \dchithreegr & \dchithreegrall   &
\dchithreelogb  & \\  
&$\delta\hat{\varphi}_4$  & $f^{-1/3}$ &  \dchifourmedian &\dchifourmedianall    &\dchifourgr & \dchifourgrall   & \dchifourlogb & \\  
&$\delta\hat{\varphi}_{5l}$  & $\log(f)$ & \dchifivelmedian &\dchifivelmedianall      &\dchifivelgr & \dchifivelgrall   & \dchifivellogb & \\  
&$\delta\hat{\varphi}_6$  & $f^{1/3}$ &  \dchisixmedian & \dchisixmedianall     & \dchisixgr & \dchisixgrall   &  \dchisixlogb & \\  
&$\delta\hat{\varphi}_{6l}$  & $f^{1/3}\log(f)$ &   \dchisixlmedian & \dchisixlmedianall      & \dchisixlgr & \dchisixlgrall   &  \dchisixllogb &  \\  
&$\delta\hat{\varphi}_7$  & $f^{2/3}$ &\dchisevenmedian
&\dchisevenmedianall &\dchisevengr & \dchisevengrall   & \dchisevenlogb & \\  
\hline
\hline
\multirow{2}{*}{intermediate regime}& $\delta\hat{\beta}_2$  &
$\log f$ & \dbetatwomedian &\dbetatwomedianall     &  \dbetatwogr &
\dbetatwograll  & \dbetatwologb &\multirow{2}{*}{\dbetaslogb} \\
& $\delta\hat{\beta}_3$  & $f^{-3}$ &
\dbetathreemedian & \dbetathreemedianall    &\dbetathreegr & \dbetathreegrall  & \dbetathreelogb &  \\ 
\hline
\multirow{3}{*}{merger--ringdown regime} &
$\delta\hat{\alpha}_2$  & $f^{-1}$ &  \dalphatwomedian & \dalphatwomedianall      & \dalphatwogr & \dalphatwograll & \dalphatwologb & \multirow{3}{*}{\dalphaslogb}\\
& $\delta\hat{\alpha}_3$  & $f^{3/4}$ &  \dalphathreemedian & \dalphathreemedianall     &  \dalphathreegr & \dalphathreegrall & \dalphathreelogb & \\
& $\delta\hat{\alpha}_4$  & $\tan^{-1}(af + b)$ & \dalphafourmedian &
\dalphafourmedianall      &  \dalphafourgr & \dalphafourgrall &
\dalphafourlogb & \\ 
\hline
\end{tabular}
\end{table*}

For each set of testing parameters, we perform a separate
\textsc{LALInference} analysis, where in concert with the full set of
GR parameters~\cite{GW150914-PARAMESTIM} we also explore the posterior distributions for the
specified set of testing parameters. Since our testing parameters are purely
phenomenological (except the parameters that govern the PN early-inspiral stage), we choose their prior probability distributions to
be uniform and wide enough to encompass the full posterior probability
density function in the single-parameter case. In particular we set 
$\delta\hat{\varphi}_i \in $ \varphiprior; 
$\delta\hat{\beta}_i \in $ \betaprior;  
$\delta\hat{\alpha}_i \in $ \alphaprior. In all cases we obtain estimates of the physical parameters --
  e.g., masses and spins -- that are
  in agreement with those reported in Ref.~\cite{GW150914-PARAMESTIM}.

In Fig.~\ref{fig:PNbounds} we show the $90\%$ upper bounds on deviations in the (known) PN 
parameters, $\delta\hat{\varphi}_i$ with $i=0,\ldots,7$ (except for $i =
5$, which is degenerate with the reference phase), when  
varying the testing parameters one at the 
time, keeping the other parameters fixed to the GR value. As an illustration, following Ref.~\cite{YunesHughes2010}, 
we also show in Fig.~\ref{fig:PNbounds} the bounds obtained from the measured orbital-period 
derivative $\dot{P}_{\rm orb}$ of the double pulsar 
J0737-3039~\cite{Wex:2014nva}. Also for the latter, bounds are computed by 
allowing for possible violations of GR at different powers of frequency, one at a time.
Not surprisingly, 
since in binary pulsars the orbital period changes at essentially a constant rate, the corresponding 
bounds quickly become rather loose as the PN order is increased. As a consequence, the double-pulsar 
bounds are significantly less informative than \TheEvent{}, except at 0PN 
order, where the double-pulsar bound is better thanks to the 
  long
observation time  ($\sim 10$ years against $\sim 0.4$ s for \TheEvent{}).\footnote{We note that when computing the upper bounds with 
the binary-pulsar observations, we include the effect of eccentricity only in the 0PN parameter. For the 
higher PN parameters, the effect is not essential considering that the bounds are not very tight.} Thus, 
GW150914 allows us for the first time to constrain the coefficients in the PN series of the phasing up to 3.5PN order.

Furthermore, in Table~\ref{tab:tiger-parameters} and Fig.~\ref{fig:summary} we summarize 
the constraints on each testing parameter $\delta \hat{\varphi}_i$ for the single and 
multiple-parameter analyses. In particular, in the $6^{\rm th}$ and $7^{\rm th}$ columns of Table~\ref{tab:tiger-parameters} 
we list the quantile at which the GR value of zero is found within the marginalized one-dimensional posterior 
(i.e., the integral of the posterior from the lower bound of the prior up to zero).  
We note that in the single-parameter analysis, for several  
parameters, the GR value is found at quantiles close to an equivalent
of $2\mbox{--}2.5\sigma$, 
i.e., close to the tails of their posterior probability functions. 
It is not surprising that this should happen for the majority of the early-inspiral parameters since we find that these 
parameters have a substantial degree of correlation. 
Thus, if a particular noise realization causes the 
posterior distribution of one parameter to be off-centered with respect to zero, we expect that the posteriors of all 
the other parameters will also be off-centered. This is indeed what we observe. The medians of the 
early-inspiral single-parameter posteriors reported in Table~\ref{tab:tiger-parameters} show 
opposite sign shifts that follow closely the sign
pattern found in the PN series.

We repeated our single-parameter analysis on $20$ datasets
obtained by adding the same NR waveform with \TheEvent{}-like parameters to
different noise-only data segments close to \TheEvent{}. 
In one instance, we
observed $\delta \hat{\varphi}_i$ posterior distributions very similar to
those of Table \ref{tab:tiger-parameters} and Fig.~\ref{fig:summary}, both in terms of 
their displacements from
zero and of their widths, whereas for the others the displacements tended
to be much smaller (though the widths were still comparable).
Thus, 
it is not unlikely that instrumental noise fluctuations would cause the 
degree of apparent deviation from GR found to occur in the single-parameter 
quantiles for GW150914, even in the absence of an actual deviation from GR.
However, we cannot fully exclude a systematic origin from inaccuracies or 
even missing physics in our waveform models. Future observations will 
shed light on this aspect. 

In the multiple-parameter analysis, which accounts for correlations between parameters, the GR value is usually found to be very close to the median of the marginalized distributions. This is partly due to the fact that we are not
  sensitive to most of the early-inspiral parameters, with
  the exception of the 0PN and 0.5PN coefficients.
As for the intermediate and merger--ringdown parameters, since most of the SNR for \TheEvent{} comes 
from the high-frequency portion of the observed signal, we find that the constraints on those coefficients 
  are very robust and essentially independent of the analysis configuration chosen, single or multiple. 

Finally, the last two columns of Table~\ref{tab:tiger-parameters} report the logarithm of the ratio
of the marginal likelihoods (the logarithm of the Bayes factor $\log_{10} B^{\rm GR}_{\rm model}$) as a measure of
the relative goodness of fit between the  \textsc{IMRPhenom} and
\textsc{gIMR} models (see Ref.~\cite{GW150914-PARAMESTIM} and references therein). If $\log_{10} B^{\rm
  GR}_{\rm model} < 0\,(> 0)$ then GR fits the data worse (better)
than the competing model. The uncertainty over $\log_{10} B^{\rm GR}_{\rm model}$ is
estimated by running several independent instances of
\textsc{LALInference}. The $\log_{10} B^{\rm  GR}_{\rm model}$ values shown in Table~\ref{tab:tiger-parameters} corroborate
our finding that \TheEvent{} provides no evidence in favor of the hypothesis that GR 
is violated.\footnote{Because of the normalization of the prior
    probability distributions, the Bayes factors include a penalty factor
  -- the so-called Occam factor -- for models that have more
  parameters. The wider the prior range for the additional parameters, the more severe the
  penalization. Therefore, different choices for $\delta\hat{p}_i$
  would lead to different numerical values of  $\log_{10}B^{\rm
    GR}_{\rm model}$. To fully establish the significance of the
    Bayes factors, validation studies~\cite{Li:2011cg,AgathosEtAl:2014} would be necessary and will be presented in forthcoming studies.}

As an aside, we note that \TheEvent{} was detected with the LIGO detectors at 
about one-third of their final design sensitivity, which is expected to be achieved 
around 2019 \cite{Aasi:2013wya}. Hence future detections are
 expected to occur with larger SNRs, leading to tighter bounds on phase
 coefficients. It is also worth noting that the posterior 
density functions for the $\delta\hat{p}_i$ from all future detections can be combined,
leading to a progressive improvement of the bounds on these parameters.

\begin{figure}[t]
\includegraphics[width=\columnwidth]{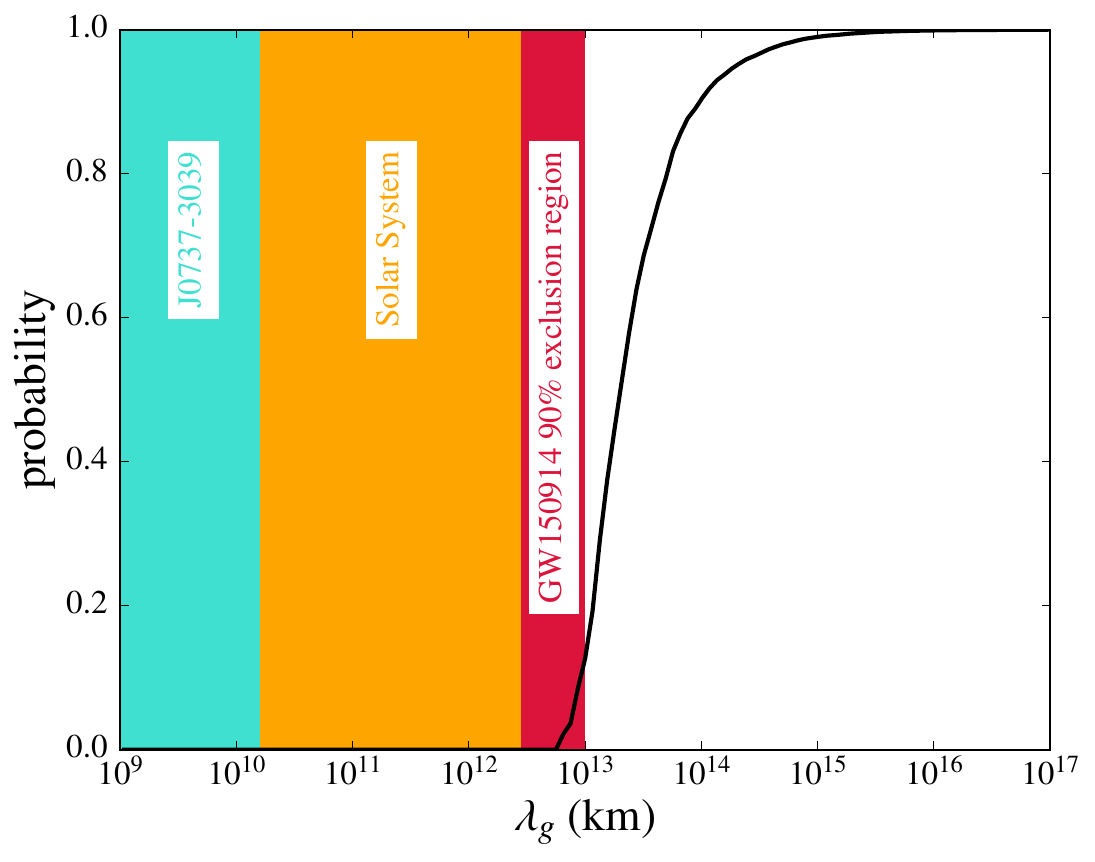}
\caption{Cumulative posterior probability distribution for $\lambda_g$ (black curve) and exclusion regions for the graviton Compton wavelength $\lambda_{g}$ 
from \TheEvent{}. The shaded areas show exclusion regions from the
double pulsar observations (turquoise), 
the static Solar System bound (orange) and the 90\% (crimson) region from \TheEvent{}.} 
\label{fig:lambdag}
\end{figure}

\paragraph{Constraining the graviton Compton wavelength.} 
Since the 1970s, there have been attempts to construct theories of gravity 
mediated by a graviton with a non-zero mass. Those attempts have led to conceptual 
difficulties; some of these have been addressed, circumvented, or overcome, but 
others remain open (see Ref.~\cite{deRham:2014zqa} and references therein). 
Here, we take a phenomenological approach and consider a hypothetical massive-graviton 
theory in which, due to a modification of the dispersion relation, 
GWs travel at a speed different from the speed of light.

In GR, gravitons are massless and travel at the speed of light $v_g = c$. 
In a massive-graviton theory the dispersion relation can be 
modified to $E^2 = p^2 c^2 + m_g^2c^4$, where $E$ is the graviton energy, $p$ 
the momentum, and $m_g$ is the 
graviton rest mass, related to the graviton's Compton wavelength by $\lambda_g=h/(m_gc)$ 
with $h$ the Planck constant. Thus, we have $v^2_g/c^2 \equiv c^2 p^2/E^2  \simeq 
1 - h^2c^2/(\lambda_g^2E^2)$, and the massive graviton propagates at an  
energy (or frequency) dependent speed. Another effect one expects on general grounds is 
that the Newtonian potential gets altered by a Yukawa-type correction whose 
characteristic length scale is $\lambda_g$:   
$\varphi(r) = (GM/r) [1 - \exp(-r/\lambda_g)]$.

Existing bounds on $\lambda_g$ that do not probe the propagation of gravitational interactions 
(i.e., the so-called {\it static} bounds), come from Solar System 
observations~\cite{Talmadge:1988qz,Will:1998} (which probe the above 
Yukawa-corrected Newtonian potential), the non-observation of superradiant 
instabilities in supermassive black holes \cite{Brito:2013wya}, 
model-dependent studies of the large-scale dynamics of galactic
clusters~\cite{GoldhaberEtal74}, and weak lensing
observations~\cite{Choudhury:2002pu}; these bounds are $2.8\times 10^{12}$ km, $2.5 \times 10^{13}$ km, 
$6.2\times 10^{19}$ km and $1.8\times 10^{22}$ km, respectively. We note that the 
bound from superradiance relies on the assumption that the very massive, compact objects 
in the centers of galaxies are indeed supermassive Kerr black holes, as opposed to other, 
more exotic objects. As also stressed in Ref.~\cite{Will:1998}, the model-dependent bounds from 
clusters and weak lensing should be taken with caution, in view of the uncertainties on 
the amount of dark matter in the Universe and its spatial distribution. 
The only {\it dynamical} bound to date comes from binary-pulsar 
observations~\cite{Finn:2001qi} and it is $\lambda_{g} > 1.6\times 10^{10}$ km.
If the Compton wavelength of gravitons is finite, then lower frequencies propagate 
slower compared to higher frequencies, and 
this dispersion of the waves can be incorporated in the gravitational phasing from a 
coalescing binary. In particular, neglecting 
all possible effects on the binary dynamics that could be introduced by the 
massive graviton theory, Ref.~\cite{Will:1998} found that 
the phase term $\Phi_{\mathrm{MG}}(f)=-(\pi D c)/[\lambda_g^2(1+z)f]$ (formally a 
1PN order term) 
should be added to the overall GW phase. In
this expression, $z$ is the cosmological redshift and $D$ is a cosmological 
distance defined 
in Eq.~(2.5) of Ref.~\cite{Will:1998}. 

\TheEvent{} allows us to search for evidence of dispersion as the
signal propagated toward the Earth. We perform the analysis by
explicitly including the formally 1PN-order term
above~\cite{Will:1998,Keppel:2010qu} in the \textsc{EOBNR} and
\textsc{IMRPhenom} GW phases and treating $\lambda_{g}$ as an
additional, independent parameter~\cite{DelPozzoEtAl:2011}. We assume
a standard $\Lambda$CDM cosmology~\cite{Planck} and a uniform prior
probability on the graviton mass $m_g\in[10^{-26},10^{-16}]$ eV/c$^2$, thus the prior on $\lambda_g$ is $\propto 1/\lambda_g^2$.
In Fig.~\ref{fig:lambdag} we show the cumulative posterior
probability distribution for $\lambda_g$ obtained from combining the
results of the two waveform models (\textsc{EOBNR} and
\textsc{IMRPhenom}) following the procedure outlined in
Ref.~\cite{GW150914-PARAMESTIM}. We find no evidence for a finite value of
$\lambda_{g}$, and we derive a dynamical lower bound $\lambda_g >
\GRAVITONCOMPTONWAVELENGTH{}$ at 90\% confidence, which corresponds to
a graviton mass $m_g\leq 1.2\times10^{-22}$ eV/c$^2$. This bound is approximately a factor of three better than
the current Solar-System bound~\cite{Talmadge:1988qz,Will:1998}, and
$\sim$ three orders of magnitude better than the bound from
binary-pulsar observations~\cite{Finn:2001qi}, but it is less
constraining than model-dependent bounds coming from the large-scale
dynamics of galactic clusters~\cite{GoldhaberEtal74}, weak
gravitational-lensing observations~\cite{Choudhury:2002pu}, and the non-observation of 
superradiant instability in supermassive black holes~\cite{Brito:2013wya}.

\paragraph{No constraint on non-GR polarization states.}
GR predicts the existence of two transverse-traceless tensor polarizations for
GWs. More general metric theories of gravitation allow for up to four
additional polarization states: a transverse scalar mode and 
three longitudinal modes \cite{Eardley:1973br,Will:2014kxa}.  
Because the Hanford and Livingston LIGO instruments have similar orientations,
they are sensitive to a very similar linear combination of the GW polarizations, so
it is difficult to distinguish between the GR and non-GR states.

As an illustration, we use the \textsc{BayesWave} GW-transient
analysis algorithm \cite{Cornish:2014kda} to reconstruct the
\TheEvent{} waveform, assuming the simplest case in which the
signal consists entirely of the transverse scalar
(breathing) mode. We compare the reconstructed waveforms and power spectral densities 
(PSDs) for the
pure scalar-mode and GR models, and find the $\log$ Bayes factor between the
two hypotheses to be $\log B^\mathrm{GR}_{\mathrm{scalar}} = 1.3 \pm 0.5$ when using the 
PSD from the breathing mode analysis and
$\log B^\mathrm{GR}_\mathrm{scalar} = -0.2 \pm 0.5$ when using the PSD from the GR analysis. In 
both cases the log Bayes factors do not 
significantly favor one model over the other. The only notable
difference is in the reconstructed sky locations; the latter reflects
the different response of the detector network to the tensor
components compared to the purely scalar mode.

We reiterate that this test is only meant to illustrate the difficulty in distinguishing between GR and non-GR polarization states on the basis of GW150914 data alone. Furthermore, the results are not in contradiction with the comprehensive parameter estimation studies of GW150914~\cite{GW150914-PARAMESTIM}, which model only the transverse-traceless GR polarizations. Finally, we note that in the weakly dynamical regime, binary pulsars~\cite{Wex:2014nva} do provide evidence in favor of GR, in that they would have a different decay rate if scalar radiation dominated. To directly study the polarization content of gravitational radiation from the strong-field dynamics, a larger network including detectors with different orientations, such as Advanced Virgo \cite{TheVirgo:2014hva}, KAGRA \cite{Aso:2013eba}, or LIGO-India~\cite{Indigo} will be required, at least in the context of unmodeled GW-signal reconstruction.

\paragraph{Outlook.} 
The observation of \TheEvent{} has given us the
opportunity to perform quantitative tests of the genuinely
strong-field dynamics of GR. We investigated the nature of \TheEvent{}
by performing a series of tests devised to detect inconsistencies
with the predictions of GR. With the exception of the graviton Compton
wavelength and the test for the presence of a non-GR polarization, we did not perform any
study aimed at constraining parameters that might arise from specific 
alternative theories \cite{Will:2014kxa,Berti:2015itd,Yunes:2013dva}, such as Einstein-{\ae}ther theory~\cite{Jacobson:2000xp} 
and dynamical Chern--Simons~\cite{Alexander:2009tp}, or from compact-object binaries composed of exotic objects such as 
boson stars~\cite{Liebling:2012fv} or gravastars~\cite{Mazur:2004fk}. Studies of this kind are not possible yet, 
since we lack predictions for what the inspiral--merger--ringdown GW signal should look like in those cases. We hope that the
observation of \TheEvent{} will boost the development of such models
in the near future.

In future work we will also attempt to measure more than one damped sinusoid from the data after \TheEvent{}'s peak, thus extracting
the QNMs and inferring the final black hole's mass and spin. We will, thus, be able to test the no-hair
theorem~\cite{Israel:1967wq,Carter:1971zc} and the second law of black-hole dynamics~\cite{Hawking:1971tu,Bardeen:1973gs}. However, signals
louder than \TheEvent{} might be needed to achieve these goals. GR predicts the existence of only two transverse polarizations for GWs. 
We plan to investigate whether an extended detector network will allow the measurement of non-transverse components~\cite{Will:2014kxa} in further GW signals.

The constraints provided by GW150914 on deviations from GR are
unprecedented due to the nature of the source, but they do not
reach high precision for some types of deviation, particularly
those affecting the inspiral regime. A much higher
SNR and longer signals are necessary for more stringent tests. However, it is not clear up to which SNR our parameterized waveform 
models are still a faithful representation of solutions of Einstein's equations. Furthermore, to extract 
specific physical effects we need waveform models that are expressed in terms of relevant parameters. 
We hope that, encouraged by \TheEvent{}, further
efforts will be made to develop reliable, physically relevant, and computationally
fast waveform models. More stringent bounds can be
obtained by combining results from multiple GW observations~\cite{DelPozzoEtAl:2011,Li:2011cg,AgathosEtAl:2014,Ghosh:2015xx}. Given the rate of coalescence of binary black holes as inferred in Ref.~\cite{GW150914-RATES}, we are looking forward
to the upcoming joint observing runs of LIGO and Virgo.

The detection of \TheEvent{} ushers in a new era in the field of
experimental tests of GR. The first result of this era is that, within the limits set by our sensitivity, all the tests performed on \TheEvent{} provided no evidence for disagreement with the predictions of GR.

\paragraph{Note.}
This version incorporates the corrections ~\cite{GW150914-TGR-Erratum}.

\paragraph{Acknowledgments.}
The authors gratefully acknowledge the support of the United States
National Science Foundation (NSF) for the construction and operation of the
LIGO Laboratory and Advanced LIGO as well as the Science and Technology Facilities Council (STFC) of the
United Kingdom, the Max-Planck-Society (MPS), and the State of
Niedersachsen/Germany for support of the construction of Advanced LIGO 
and construction and operation of the GEO600 detector. 
Additional support for Advanced LIGO was provided by the Australian Research Council.
The authors gratefully acknowledge the Italian Istituto Nazionale di Fisica Nucleare (INFN),  
the French Centre National de la Recherche Scientifique (CNRS) and
the Foundation for Fundamental Research on Matter supported by the Netherlands Organisation for Scientific Research, 
for the construction and operation of the Virgo detector
and the creation and support  of the EGO consortium. 
The authors also gratefully acknowledge research support from these agencies as well as by 
the Council of Scientific and Industrial Research of India, 
Department of Science and Technology, India,
Science \& Engineering Research Board (SERB), India,
Ministry of Human Resource Development, India,
the Spanish Ministerio de Econom\'ia y Competitividad,
the Conselleria d'Economia i Competitivitat and Conselleria d'Educaci\'o, Cultura i Universitats of the Govern de les Illes Balears,
the National Science Centre of Poland,
the European Commission,
the Royal Society, 
the Scottish Funding Council, 
the Scottish Universities Physics Alliance, 
the Hungarian Scientific Research Fund (OTKA),
the Lyon Institute of Origins (LIO),
the National Research Foundation of Korea,
Industry Canada and the Province of Ontario through the Ministry of Economic Development and Innovation, 
the Natural Science and Engineering Research Council Canada,
Canadian Institute for Advanced Research,
the Brazilian Ministry of Science, Technology, and Innovation,
Russian Foundation for Basic Research,
the Leverhulme Trust, 
the Research Corporation, 
Ministry of Science and Technology (MOST), Taiwan
and
the Kavli Foundation.
The authors gratefully acknowledge the support of the NSF, STFC, MPS, INFN, CNRS and the
State of Niedersachsen/Germany for provision of computational resources.

Finally, we thank the anonymous referees, whose comments helped improve the clarity 
of the paper.

\def\url{} 

\bibliographystyle{apsrev-nourl}
\bibliography{references,GW150914_refs}

\end{document}